\documentstyle[onecolumn]{mn}

   \def\lesssim{\mathrel{\hbox{\rlap{\hbox{\lower4pt\hbox{$\sim$}}}\hbox{$<$}}}}
   \def\gtrsim{\mathrel{\hbox{\rlap{\hbox{\lower4pt\hbox{$\sim$}}}\hbox{$>$}}}}
   
   \newcommand{\bm}[1]{\mbox{\boldmath$#1$}}
   \newcommand{\mal}[1]{\stackrel{_\circ}{#1}}
   \newcommand{\kaco}[1]{\left\langle{#1}\right\rangle}
   \makeatletter
\renewcommand{\theequation}{%
       \thesection.\arabic{equation}}
\@addtoreset{equation}{section}
\makeatother
   \def\theenumi{\arabic{enumi}}

\begin{document}

\title[PN Lagrangian Perturbation Approach in the (3+1) formalism]
{\bf Post-Newtonian Lagrangian Perturbation Approach to the Large-Scale 
Structure Formation}

\author[M. Takada and T. Futamase]
{Masahiro Takada\footnotemark[1]  and    
Toshifumi Futamase\footnotemark[2] \\
\sl Institute of Astronomy, Graduate School of Science,
Tohoku University, Sendai 980-8578, Japan}

\footnotetext[1]{Electronic address: takada@astr.tohoku.ac.jp}
\footnotetext[2]{Electronic address: tof@astr.tohoku.ac.jp}

\maketitle

\begin{abstract}
We  formulate the Lagrangian perturbation theory to solve 
the non-linear dynamics of self-gravitating  fluid  within 
the framework of the post-Newtonian approximation in general relativity, 
using the (3+1) formalism. 
Our formulation coincides with 
Newtonian Lagrangian perturbation theory developed by Buchert for the
scale 
much smaller than the horizon scale, and
with the gauge invariant linearized theory in 
 the longitudinal gauge conditions  for the linear regime. 
These are  achieved by using the gauge invariant quantities 
at the initial time when the linearized theory is valid enough. 
The post-Newtonian corrections in the solution of the trajectory field of 
fluid elements are calculated in the explicit forms. 
Thus our formulation allows us to investigate the evolution 
of the large-scale fluctuations involving relativistic corrections
from the early regime  such as the decoupling time of matter and
 radiation 
 until today.
As a result, we are able to show that naive Newtonian cosmology to
 the structure formation 
will be a good approximation even  for the perturbations with scales 
not only inside 
 but also  beyond the present horizon scale in the longitudinal coordinates.
Although the post-Newtonian corrections are small, it is shown that 
they  have a growing transverse mode which is not   present 
 in Newtonian theory as well as  in the gauge invariant linearized theory. 
Such post-Newtonian order effects might produce characteristic 
appearances of the large-scale structure formation, for example, 
through the observation of anisotropies of the cosmic microwave 
background  radiation (CMB).
Furthermore  since our approach has a straightforward Newtonian limit,
it  will be also convenient for numerical implementation based 
on the presently available Newtonian simulation.
Our results easily allow us to perform a 
simple order estimation of each term 
in the solution, which  indicates that 
the post-Newtonian corrections may not be neglected in the early evolution 
of the density fluctuation compared with Newtonian perturbation
 solutions. 
\end{abstract}


\section{Introduction}

The observation of isotropy of the cosmic microwave background (CMB) 
indicates that the universe is remarkably isotropic over the horizon
scale \cite{smoot}. 
Thus it is natural to describe the large scale spatial geometry
of the universe by a homogeneous and isotropic metric, namely, the
Friedmann-Robertson-Walker (FRW) model. 
However, the real universe is neither isotropic nor homogeneous on  local
scales and has a hierarchical structure such as  galaxies,
clusters of galaxies, superclusters of galaxies and so on.  
In the standard big bang scenario, it is commonly believed that
such   large-scale inhomogeneous structures of the universe were formed
as
a result of growth of density fluctuations whose amplitudes 
had been very small in the early regime in FRW background 
(see, e.g., Peebles \shortcite{peebles}). 
Furthermore, such inhomogeneous structures are
usually considered to have been formed only 
due to gravitational instability
 after matter (baryons) and radiation decoupled.
Therefore, the gravitational instability of self-gravitating fluid  
has been a main subject of the research in cosmology in connection to
the formation of structures.

The investigation of the nonlinear large-scale structure formation of
the universe have been usually studied within the Newtonian theory. 
This is based on the following fact: 
as far as the scales of the considered density fluctuations 
are much smaller than the horizon scale, the Newtonian approximation 
will be accurate enough to describe their evolutions up to non-linear 
regime where the density contrast  becomes larger than unity \cite{peebles}.
Actually, many large-scale simulations based on the Newtonian approximation 
have been performed 
to compare the numerical results with the observed distributions of 
galaxies. Analytical approaches to the evolution of density fluctuation 
based on the Newtonian Lagrangian picture have also been developed 
by Buchert \shortcite{buchert89,buchert92,buchert94} and the other authors 
\cite{bouchet92,bouchet,catelan,moutarde} where it is expected that the 
approach gives a good approximation up to a certain stage of non-linear 
regime \cite{buchertW,mbw,weiss}. Buchert 
\shortcite{buchert92,buchert93} 
showed that the general first-order
solutions for an Einstein-de Sitter background include well-known
Zel'dovich approximation \cite{z} as a special case.  

On the other hand, the region of observations of the large-scale 
structure is steadily increasing. For example,the very ambitious 
observational program Sloan Digital Sky
Survey \cite{SDSS} will complete a spectroscopic survey of 
$\sim 10^6$ galaxies brighter than $r'\sim18$ mag 
over $\pi$ steradians, which covers 
the region of several hundred megaparsecs.
As mentioned above,  
Newtonian approximation is valid only for system
with scales much smaller than the horizon scale, so
 it is not clear at all if the application of the Newtonian 
theory is appropriate for  such a wide region of spacetime.
In fact the fluctuations relevant for the large-scale structures 
have not been always  much smaller than horizon scales in the past. 
For example, the horizon scale at the decoupling
time is of the order of $ct_{\mbox{{\scriptsize dec}}}
(1+z_{\mbox{{\scriptsize dec}}})\sim
80h^{-1}\mbox{Mpc}$ in the present physical length. 
This suggests that we have to employ the relativistic description 
for the evolution of fluctuations larger than or equivalent to
 such scales.

Thus to understand the evolution of the large-scale structure of the
universe closely, 
it is important and necessary to have
some formalism to evaluate  the effect of general
relativistic corrections to the  Newtonian dynamics on such a region. 
Conversely, until we obtain such a  formalism,  
we are not confident about  the validity of the Newtonian cosmology 
in  studying evolutions of perturbations. 
The purpose of the present  paper is to provide such a formalism based on
the Lagrangian perturbation theory in the post-Newtonian (PN) framework. 
In  the PN approximation to cosmology \cite{futamase96,sa},
it has been shown  that the metric is given in the following form at the
Newtonian limit 
\[
 ds^2=-\left(1+2\frac{\phi}{c^2}\right)(cdt)^2+a^2\!(t)\left(
1-2\frac{\phi}{c^2}\right)dl^2, 
\]
and the Newtonian-like 
gravitational potential $\phi$ 
is generated  by the density fluctuation field $\delta$
via the cosmological Poisson equation
$\Delta\phi=4\pi G a^2\rho_b\delta$, where $\rho_b$
denotes the background density. This includes the non-linear 
situations for the density contrast field where is much larger 
than unity.  
%%%%%%%%%%%%%%%%%%%%%%%%%%%%%%%%%%%%%%%%%%%%%%%%%%%%%%%%%%%%%%%%%%%%%%
The above  metric is also applied in 
deriving the cosmological lens equation 
for a realistic inhomogeneous universe \cite{futamaselens,pyne}.  
For recent years,
the studies of the gravitational lensing in the cosmological
circumstance have been widely developed using the equation. 
For example, Seljak \shortcite{seljak} investigated
the secondary effect of gravitational lensing on 
 CMB anisotropies induced by  
 the non-linear large-scale structures using the above
{\it Newtonian} metric (longitudinal gauge). 
The future generation of detectors of CMB anisotropies 
with great accuracy requires theoretical efforts to obtain 
more precise predictions for the anisotropies. 
In such a case  
 we need to consider the gravitational lensing 
of photons traveling across the 
present horizon scale ($\sim 3000h^{-1}\mbox{Mpc}$).  
This  
indicates that, if using the above metric, 
we first have to investigate the validity whether the cosmological 
Newtonian metric is able to 
 describe accurately the dynamics of fluctuations 
within such a scale. 
%Usually, it seems to be naively regarded  that Newtonian 
%cosmology is a good approximation in a realistic universe. 
%Here, we will make clear the problem rigorously. 
The following considerations may explain our results of the present 
paper intuitively.   
%%%%%%%%%%%%%%%%%%%%%%%%%%%%%%%%%%%%%%%%%%%%%%%%%%%%%%%%%%%%%%%%%%%%%%
The cosmological Poisson equation 
 suggests that the metric perturbation quantity $\phi/c^2$ is 
of the order of $(al/(ct))^2\delta_{(l)}$ for the density contrast
$\delta_{(l)}$
with the characteristic scale $l$ in the present physical length. 
Then we expect that the  first PN metric perturbations  
which determine the first PN  force of
equations of motion are  of the order of $(\phi/c^2)^2$, 
namely, $(al/(ct))^4\delta^2_{(l)}$. 
This  means that the quantity $(al/(ct))^2\delta_{(l)}$ is
an essential expansion parameter in the  PN approach to cosmology. 
As mentioned later in detail,
the fact that the quantity $(l/(ct_0))^2\delta(t_0)$ is much smaller 
than unity for the present horizon-scale fluctuation with the usual
power spectra explains why the PN approximation is sufficiently 
 able to
describe the evolution of the fluctuations not
only inside but also beyond  the present horizon scale $ct_0$. 
Furthermore, the formalism based on the PN approximation 
allows us, for example,  a more accurate treatment 
of propagation of photon on the geometry produced by matter
inhomogeneities, as required in the study of gravitational lensing 
by cosmic structures (see, e.g., Schneider, Ehlers, \& Falco1992) and other applications. 

Naturally, we also require our formalism to agree with  the gauge invariant
linearized theory developed by Bardeen \shortcite{bardeen} and Kodama \&
Sasaki \shortcite{ks} in the linear regime.  
 There have been some studies to investigate the nonlinear large-scale  
structure formation based on the relativistic Lagrangian
perturbation theory \cite{kasai93,kasai95,mps2,mt}. 
However, because of the gauge condition adopted in these studies, 
namely,  the synchronous comoving coordinates, it is not easy to 
have contact with the Newtonian Lagrangian approach fully developed 
and used for the numerical simulation. 
%%%%%%%%%%%%%%%%%%%%%%%%%%%%%%%%%%%%%%%%%%%%%%%%%%%%%%%%%%%%%%%%%%%%%%
In particular, Matarrese \& Terranova \shortcite{mt} called their 
formalism post-Newtonian treatment in spite of starting with the 
synchronous comoving coordinates. As is well  known, a peculiar velocity 
of any fluid element does not appear explicitly in such a coordinate,
which makes the PN treatment difficult.  
%%%%%%%%%%%%%%%%%%%%%%%%%%%%%%%%%%%%%%%%%%%%%%%%%%%%%%%%%%%%%%%%%%%%%
Thus we will not restrict ourselves to such a coordinates, but we shall use 
(3+1) formalism of general relativity to admit various choices of the 
coordinates. In particular we shall study the Lagrangian perturbation
theory
in the coordinates where the comparison with the  Newtonian case is 
most easily made. In this way our formalism would be also convenient to
use for the numerical simulation straightforwardly based on Newtonian
simulations. 

There have been  another relativistic Lagrangian approaches. 
Bertschinger \& Hamilton \shortcite{berthamil} and Bertschinger \&
Jain \shortcite{bertjain}  have focused on the 
Lagrangian evolution equations  of the electric and magnetic parts of
the Weyl tensor for cold dust in the Newtonian limit. They derived the 
equations using not Einstein equations but 
 the relativistic kinematical prescription for fluid flow 
 developed by 
Ehlers \shortcite{Ehlers} and Ellis \shortcite{Ellis}. They suggested 
that the magnetic part does not vanish generally in the Newtonian 
limit, implying that the Lagrangian evolution of the fluid is not 
purely local. This is a so-called ``non-local'' problem also discussed by 
the other authors \cite{hui,kofman}. Here, 
we formulate  another relativistic 
Lagrangian perturbation theory  for fluid motion from different point  
of  view.      

%%%%%%%%%%%%%%%%%%%%%%%%%%%%%%%%%%%%%%%%%%%%%%%%%%%%%%%%%%%%%%%
Recently, Matarrese, Mollerach, \& Bruni \shortcite{mpphy} 
derived the second order solutions of 
metric perturbations in the {\it Poisson} gauge that 
 reduces to the longitudinal gauge 
in the case 
of scalar first order perturbations.   
 They obtained the solutions by 
 transforming the known 
 solutions in the synchronous comoving coordinates 
to them in the Poisson gauge, using the second order gauge 
transformation. We here  start with the (3+1) formalism which includes 
 the Poisson gauge from the beginning . Therefore,  
our PN treatment may be more transparent physically, and 
can 
 give a physical understanding of 
 the validity of the Newtonian treatment.  
Actually, the time dependence of 
our second order solutions agrees with their results as described 
below.  
%%%%%%%%%%%%%%%%%%%%%%%%%%%%%%%%%%%%%%%%%%%%%%%%%%%%%%%%%%%%

This paper is organized as follows. 
In Section 2 we shall write down the cosmological PN
approximation in (3+1) formalism. In Section 3 we introduce the 
Lagrangian perturbation theory to iteratively solve the PN
equations derived in Section 2. In Section 4  we summarize our results
and discuss 
the effect of PN corrections and evaluate explicitly its
importance. Summary and conclusions are given in Section 5. 
Throughout this paper,
Greek and Latin indices take $0,1,2,3$ and $1,2,3$, respectively.

\section{The post-Newtonian approximation in (3+1) formalism}

\subsection{The (3+1) formalism in cosmological situation}

We wish to develop the relativistic 
Lagrangian perturbation theory 
formulated in the coordinates system which has a 
straightforward 
Newtonian limit. 
For this purpose, it is convenient 
to use the (3+1) formalism. The post-Newtonian (PN) approximation 
in the (3+1) formalism has been formulated by one of authors \cite{futamase92} 
and further developed by Asada, Shibata \& Futamase \shortcite{asf}.
Its cosmological generalization is straightforward \cite{futamase96,sa}. 
We first review the (3+1) formalism in 
the cosmological situation. 

Here we shall present the basic equations using the (3+1) formalism.
Let us first assume that there exists a congruence of timelike
worldlines from which the spacetime looks isotropic. We shall call the
family of the worldlines basic observers, who see no dipole component
 of anisotropies of 
the cosmic microwave background (CMB). We can regard  that  any one of
these observers at the spacetime point $x$ moves with 4-velocity
$n^\mu(x)$
 without loss of
generality. The tangent vector $n^\mu$ is  
normalized as $n^\mu
n_\mu=-1$. These observers are used to foliate the spacetime by their
simultaneous surfaces: $t=\mbox{const.}$ We shall assume that the three
surfaces form hypersurfaces with the unit normal vector field $n^\mu$.

In the (3+1) formalism, the metric is split as 
\begin{eqnarray}
& &g_{\mu \nu}=\gamma_{\mu\nu}-n_\mu n_\nu, \nonumber \\
& &n_{\mu}=( -\alpha ,\mbox{{\bf 0}} ), \nonumber  \\
& &n^{\mu}=\left( \frac{1}{\alpha},-\frac{\beta ^{i}}{\alpha} \right), 
\end{eqnarray}
where $\gamma_{\mu\nu}$ is the projection tensor and
$\alpha$, $\beta^i$, and $\gamma_{ij}$ are the lapse function, shift vector and metric
 on the spatial hypersurface, respectively. 
Then the line element may be written as 
\begin{eqnarray}
ds^2&=& -\left(n_\mu dx^\mu\right)^2+\gamma_{\mu \nu}dx^\mu dx^\nu
 \nonumber \\
 &=&-(\alpha^2-\beta_{i} \beta^{i})(cdt)^2+2\beta_{i}(cdt)dx^{i}
+\gamma_{ij}dx^{i}dx^{j}
\label{eqn:aa}
\end{eqnarray}
We shall keep the power of $c$ explicitly henceforth 
because we make the $1/c$ expansion in various quantities in 
the post-Newtonian approximation. 

We may define the extrinsic curvature of the hypersurface by
\begin{eqnarray}
&&K_{ij} = -  n_{i;j}
\end{eqnarray}
where $;$ denotes the covariant derivative  with respect to the  
metric  $g_{\mu\nu}$. The basic variables in the (3+1) formalism 
are then the spatial metric 
and the extrinsic curvature. The lapse function and shift vectors are 
freely specified as the coordinate conditions. 
The (3+1) formalism naturally decomposes the Einstein equations,
\[
G_{\mu\nu}=\frac{8\pi G}{c^4}T_{\mu\nu}-\Lambda g_{\mu\nu},
\]
into the constraint equations and the evolution equations. 
$\Lambda$ denotes the cosmological constant.
The former set of equations constitutes the so-called
 Hamiltonian and momentum constraints. These become
\begin{eqnarray}
&&R-K_{ij}K^{ij}+K^2=\frac{16\pi G}{c^4}E+2\Lambda, \label{eqn:ab}\\
&&D_{i}K^{i}{}_{j}-D_{j}K=\frac{8\pi G}{c^4}J_{j}, \label{eqn:ac}
\end{eqnarray}
where  $K = \gamma^{ij}K_{ij}$, $D_i$ and $R$ are the covariant derivative
and  the  scalar curvature with respect to $\gamma_{ij}$, respectively.   
$E$ and $J_{i}$ are defined as
\begin{eqnarray}
&&E=T_{\mu\nu}n^{\mu}n^{\nu}, \nonumber \\
&&J_{i}=-T_{\mu\nu}n^{\mu}\gamma^{\nu}{}_{i}. \label{eqn:ad}
\end{eqnarray}
The evolution equations for the spatial metric and the 
extrinsic curvature become
\begin{eqnarray}
&&\frac{1}{c}\frac{\partial }{\partial t} \gamma_{ij} = -2 \alpha K_{ij}+D_{i} \beta_{j}
+D_{j} \beta_{i},  \label{eqn:ae}\\
&&\frac{1}{c}\frac{\partial }{\partial t}K_{ij} = \alpha (R_{ij}+KK_{ij}-2K_{il}K^{l}{}_{j})
-D_{i}D_{j} \alpha -\alpha \Lambda \gamma_{ij} \nonumber \\
& &\hspace{5em} +(D_{j} \beta^{m})K_{mi}+(D_{i}\beta ^{m})K_{mj}+\beta ^{m}D_{m}K_{ij}
-\frac{8\pi G}{c^4}\alpha \left( S_{ij}+\frac{1}{2} \gamma _{ij}(E-S^{l}{}_{l})\right),
\label{eqn:af}
\end{eqnarray}
where $R_{ij}$ is the Ricci tensor of the hypersurface, and 
\begin{equation}
S_{ij}=T_{\mu \nu }\gamma^{\mu}{}_{i}\gamma^{\nu}{}_{j}.
\end{equation}
For the sake of convenience we shall write down the evolution equations for 
the trace of the spatial metric and the extrinsic curvature which 
will be sometimes used in the following:
\begin{eqnarray}
&&\frac{1}{c}\frac{\partial}{\partial t}\gamma =2 \gamma (-\alpha K+D_{i}\beta ^i),
\label{eqn:ag}\\
&&\frac{1}{c}\frac{\partial}{\partial t}K =\alpha (R+K^2)-D^{i}D_{i}\alpha 
+\beta ^{j}D_{j}K+\frac{4 \pi G}{c^4}\alpha (S^{l}{}_{l}-3E)
-3 \alpha \Lambda, \label{eqn:ah}
\end{eqnarray}
where $\gamma$ is the determinant of $\gamma_{ij}$.

For the cosmological application we introduce 
the scale factor which expresses the expansion of the universe. 
Following  Ehlers \shortcite{Ehlers}, 
we define the spatial distance between two nearby timelike worldlines 
with tangent vector $n^\mu$ as
\begin{eqnarray}
&&\delta \ell = (\gamma_{\mu\nu} \eta^\mu \eta^\nu )^{1/2},
\end{eqnarray}
where $\eta^\mu$ is the vector connecting these two worldlines. Then 
the change of the distance along the worldline may be calculated as 
\begin{eqnarray}
&&\frac{d}{d\tau}(\delta \ell) = n^\mu (\delta \ell)_{;\mu} 
= - [\, {1\over 3}K + \sigma_{ij} e^i e^j ] \delta \ell
\label{exp}
\end{eqnarray}
where $e^i = \eta^i/\delta \ell$, with $\eta^i = \gamma^i{}_\mu \eta^\mu$, is the 
unit spatial vector and $\sigma_{ij}$ is the trace-free part of the extrinsic 
curvature. 
It is natural to interpret the above equation (\ref{exp}) as 
expressing the {\it cosmic} expansion, that is,  
the first and the second terms 
on the right-hand side represent the isotropic and anisotropic expansion,
respectively. Motivated by this interpretation, we shall introduce the scale factor as 
\begin{eqnarray}
&&K = - {3\over c} {{\dot a}\over a},
\end{eqnarray}
where
\[
\dot{a}(t)=\left.\frac{\partial a}{\partial t}\right|_x=\frac{d a}{d t},
\]
where $a= a(t)$ is the scale factor as a  function of the cosmic time $t$. 
This gives a slice condition for $\alpha$ corresponding to the maximal 
slicing in an asymptotically flat case. For more general slice conditions, 
we shall take the following form for $K$:
\begin{eqnarray}
&&K = - {3\over c} {{\dot a}\over a} + \theta(x). \label{eqn:extrinsic}
\end{eqnarray}

It turns out that it is convenient to use the following variables 
instead of the original variables:
\begin{equation}
\tilde{\gamma}_{ij}=a^{-2}(t)\gamma_{ij},\hspace{2em} \tilde{R}=a^2 R.
\end{equation}
We also define $ \tilde{\sigma}_{ij}$ as
\begin{equation}
\tilde{\sigma}_{ij}\equiv a^{-2}\sigma_{ij}= a^{-2} \left( K_{ij}
-\frac{1}{3}\gamma _{ij}K \right).
\end{equation}
We should note that in our notation,
indices of $\tilde{\sigma}_{ij}$ are raised and 
lowered by $\tilde {\gamma }_{ij}$, so that the relations 
$\tilde{\sigma}^i\,\!_j= \sigma^i\,\!_j$ 
and $ \tilde{\sigma}^{ij}=a^2\sigma ^{ij}$ hold. 
Using these variables, the constraint equations may be rewritten as
\begin{eqnarray}
& &\frac{1}{c^2}\frac{\dot{a}^2}{a^2}+\frac{1}{6a^2}\tilde{R}-\frac{1}{6}
\tilde{\sigma}_{ij}\tilde{\sigma}^{ij}-\frac{2}{3c}\frac{\dot{a}}{a}\theta
+\frac{1}{9}\theta^2=\frac{8\pi G}{3c^4}E+\frac{1}{3}\Lambda, \label{eqn:an}\\
& & \tilde{D}_i \tilde{\sigma}^i{}_j-\frac{2}{3}\tilde{D}_j \theta=\frac{8 \pi G}{c^4}J_j,
\label{eqn:ao} 
\end{eqnarray}
where $\tilde{D}_i$ and $\tilde{R}$ are the covariant derivative 
and the scalar curvature with respect to $\tilde{\gamma }_{ij}$, respectively. 
The evolution equations take the following forms:
\begin{eqnarray} 
& &\frac{1}{c} \frac{\partial}{\partial t} \tilde{\gamma }_{ij} =\frac{2}{c}(\alpha-1)
\frac{\dot{a}}{a}\tilde{\gamma}_{ij} 
 -2\alpha \tilde{\sigma}_{ij}-
\frac{2}{3}\alpha\theta\tilde{\gamma}_{ij}+ \frac{1}{a^2}
\left( \tilde{D}_i\beta_j+\tilde{D}_j\beta_i \right), \label{eqn:aj}\\
& &\frac{1}{c} \frac{\partial}{\partial
 t}\tilde{\sigma}_{ij}=\frac{1}{a^2}
\left[ \alpha 
\left(R_{ij} 
-\frac{1}{3}\gamma _{ij}R \right) -\left( D_{i}D_j \alpha -\frac{1}{3}\gamma _{ij}
D^kD_k \alpha \right)\right]-\frac{1}{c}(\alpha+2)\frac{\dot{a}}{a}
\tilde{\sigma}_{ij}+\frac{\alpha}{3}\theta
\tilde{\sigma}_{ij} \nonumber \\
& &\hspace{5em}-2 \alpha \tilde{\sigma}_{il}\tilde{\sigma}^l{}_j+
(\tilde{D}_i\beta^l)\tilde{\sigma}_{lj}+(\tilde{D}_j\beta^l)\tilde{\sigma}_{li}
+\beta^l(\tilde{D}_l\tilde{\sigma}_{ij})-\frac{8 \pi G}{c^4 a^2}\alpha 
\left(S_{ij}-\frac{1}{3}\gamma_{ij}S^{l}\,\!_{l}
\right), \label{eqn:ak}\\
& &\frac{1}{c}\frac{1}{\tilde{\gamma}}\frac{\partial}{\partial t}
\tilde{\gamma}=\frac{6}{c}(\alpha-1)
\frac{\dot{a}}{a}+2\left(-\alpha \theta+\tilde{D}_i\beta^i \right), \label{eqn:al}\\
& &\frac{3}{c^2}\frac{\ddot{a}}{a}+\frac{3}{c^2}(\alpha-1)\frac{\dot{a}^2}{a^2}
-\frac{1}{c}\frac{\partial \theta}{\partial t} =- \alpha 
\left( \tilde{\sigma}_{ij}\tilde{\sigma}^{ij}
-\frac{2}{c}\frac{\dot{a}}{a}\theta+\frac{\theta^2}{3}\right)
+\frac{1}{a^2}\tilde{D}^i\tilde{D}_i\alpha-\beta^i\tilde{D}_i\theta-
\frac{4\pi G }{c^4}\alpha(S^i\,\!_i+E)+\alpha \Lambda, \label{eqn:am}
\end{eqnarray}
where $\tilde{\gamma}$ denotes the determinant of $\tilde{\gamma}_{ij}$.

We shall now introduce the spatial background geometry 
motivated by the following consideration. 
It seems that the metric perturbations in the present
universe remain small almost everywhere even when the density contrast
is much larger than unity. Thus it is natural to assume that the metric
structure of the universe may be described by a small perturbation from
the Friedmann-Robertson-Walker (FRW) spacetime in an appropriate
coordinate system. 
Then,
we shall further specify the spatial metric as \cite{bardeen,ks}
\begin{equation}
\tilde{\gamma}_{ij}=(1-2 \psi )\tilde{\gamma}^{(B)}_{ij}+h_{ij},\hspace{1cm} 
\tilde{\gamma}^{(B)ij}h_{ij}=0 , \label{eqn:ap}
\end{equation}
where $\tilde{\gamma}^{(B)}$ is the spatial metric of the FRW geometry: 
\begin{equation}
\tilde{\gamma}^{(B)}_{ij}=\frac{1}{ ( 1+\frac{{\cal K}}{4}r^2)^2 }\delta_{ij},
\label{eqn:aq}
\end{equation}
where ${\cal K}$ represents the curvature parameter of FRW models and 
$r^2=x_1^2+x_2^2+x_3^2$. We treat 
the perturbations as being small quantities, 
so we are able to regard spatial tensorial quantities in
our equations as tensors with respect to the background spatial metric
$\tilde{\gamma}^{(B)}_{ij}$.

Our metric is completely general. 
To reduce the gauge freedom we impose the following 
transversality constraints on $h_{ij}$ \cite{sa}:
\begin{equation}
\tilde{\gamma}^{(B)jk}\tilde{D}^{(B)}_{k}h_{ij}=0,
\label{eqn:a}
\end{equation}
where $\tilde{D}^{(B)}_i$ denotes covariant derivative with respect
to $\tilde{\gamma}^{(B)}_{ij}$, 
whose inverse is $\tilde{\gamma}^{(B)ij}$. 
By means of the condition 
({\ref{eqn:a}}), we can erase the scalar part and the 
vector part in $h_{ij}$, and it guarantees that $h_{ij}$ only contains
the tensor mode in the PN order which represents the freedom of the
gravitational wave. Thus we need to take into account
only the linear term in $h_{ij}$ because we are interested in the first PN
approximation in the present  paper. 
If we erase  
the scalar part of the shift vector by using the residual gauge 
freedom, the coordinate condition corresponds to the 
so-called ``longitudinal gauge'' (with respect to scalar modes) 
in the linear theory 
or ``Poisson gauge'' discussed by
Bertschinger \shortcite{bertschinger} and Ma \& Bertschinger 
\shortcite{mabertschinger}. 
However, we leave the freedom 
 here for the present and adopt  only the condition
(\ref{eqn:a}) for the sake of generality. 
Then the inverse matrix of $\tilde{\gamma}_{ij}$ become 
\begin{eqnarray}
& &\tilde{\gamma}^{ij}=\frac{1}{1-2\psi}\tilde{\gamma}^{(B)ij}-h^{ij},
\end{eqnarray}
where $h^i{}_j=\tilde{\gamma}^{(B)ik}h_{kj}$.  Using the above
variables, we can rewrite $\tilde{R}_{ij}$ as
\begin{eqnarray}
\lefteqn{\tilde{R}_{ij}=2{\cal K}\tilde{\gamma}^{(B)}_{ij}+\frac{1}{1-2 
\psi}(\tilde{D}^{(B)}_{i}
\tilde{D}^{(B)}_j\psi +\tilde{\gamma}^{(B)}_{ij}\tilde{D}^{(B)k}
\tilde{D}^{(B)}_k\psi)} \nonumber \\
& &\hspace{4em}+\frac{1}{(1-2 \psi)^2}\left( 3 (\tilde{D}^{(B)}_i\psi)(\tilde{D}^{(B)}
_j\psi)+\tilde{\gamma}^{(B)}_{ij}(\tilde{D}^{(B)}_k\psi)(\tilde{D}^{(B)k}\psi) \right)
+3{\cal K} h_{ij}-\frac{1}{2}\tilde{D}^{(B)}_k\tilde{D}^{(B)k}h_{ij}.
\label{eqn:ar}
\end{eqnarray}
The above discussion is adequate even if   
the universe has the nonlinear structures.  

 As one of the present authors has pointed
out \cite{futamase89,futamase96}, the problem of the back reaction of the local
inhomogeneities on the global expansion arises when the universe
has nonlinear structures ($\delta\gg 1$) on small scales.
However,  we are here interested in the  evolutions of the fluctuations from 
sufficiently early time such as the decoupling time of matter and radiation. 
Then the linear theory will be valid in the early evolution. 
Thus 
we assume that we can define the background to obey
Friedmann law until the universe has  quasi-nonlinear structures $(\delta
\gtrsim 1)$.  
In other words, we assume that even if the back reaction for the expansion 
of background exists, it is negligibly small so that 
we can perform  the perturbative approach superimposed on FRW cosmologies. 
Accordingly, from the lowest order of Eq.(\ref{eqn:an}),
we first set  
\begin{equation}
H(t)^2=\left( \frac{\dot{a}(t)}{a(t)}\right)^2=\frac{8 \pi G \rho _b(t)}{3}
-\frac{{\cal K}c^2}{a^2}
+\frac{c^2 \Lambda}{3},\label{eqn:at}
\end{equation}
where $H(t)$ and $\rho_b(t)$ are the Hubble parameter and 
the homogeneous density of the background FRW universe, respectively. 
For simplicity, we may regard $\rho_b$ as an averaged value over 
the volume as large as the horizon scale $(ct)^3$:
$\kaco{\rho}_{(ct)^3}\equiv\rho_b(t)$.  The continuity equation 
shows that the averaged mass density is conserved, namely, 
$\rho_ba^3$ is constant for the dust model. 

In the following, we only consider the flat universe (${\cal K}=0$) and use
Cartesian coordinates for simplicity.
If we fix $\theta$, Eqs.(\ref{eqn:am}), (\ref{eqn:an}), 
and (\ref{eqn:ak}) become respectively
\begin{eqnarray}
& &\frac{1}{a^2}\tilde{D}_i\tilde{D}^i\alpha =\frac{4 \pi G}{c^4}\alpha (S^i{}_i+E)+\alpha
(\tilde{\sigma}^{ij}\tilde{\sigma}_{ij}+\frac{8 \pi G }{c^2}\rho_b)-\frac{12 \pi G }{c^2}
\rho_b-\frac{1}{c}\frac{\partial \theta}{\partial t}-2\alpha \frac{1}{c}\frac{\dot{a}}{a}
\theta +\frac{\alpha}{3}\theta^2+\beta^i\tilde{D}_i\theta, \label{eqn:au}\\
&&\Delta _f\psi+\frac{3}{2}\frac{1}{1-2\psi}\psi,_k\psi,_k= \frac{4 \pi G a^2}{c^4}
E(1-2 \psi)^2+\frac{(1-2\psi)^2a^2}{4}\left(\tilde{\sigma}_{ij}\tilde{\sigma}^{ij}
-\frac{16 \pi G }{c^2}\rho_b+\frac{4}{c}\frac{\dot{a}}{a}\theta 
-\frac{2}{3}\theta^2\right), \label{eqn:av}\\
& &\frac{1}{c}\frac{\partial}{\partial t} \tilde{\sigma }_{ij}=\frac{1}{a^2} \left[ 
-\frac{\alpha}{2}\Delta _fh_{ij}+\alpha \left( \tilde{R}^{\psi}_{ij}-\frac{1}{3}
\tilde{\gamma}_{ij}\tilde{R}^{\psi}\right)
-\left( \tilde{D}_i\tilde{D}_j\alpha -\frac{1}{3}\tilde{\gamma}_{ij}
\tilde{D}^k\tilde{D}_k\alpha\right)\right]-\frac{1}{c}( \alpha+2) \frac{\dot{a}}{a}
\tilde{\sigma}_{ij}+\alpha \left(\frac{\theta}{3}\tilde{\sigma}_{ij}-2\tilde{\sigma}_{il}
\tilde{\sigma}^l{}_j\right) \nonumber \\
&&\hspace{8em}+(\tilde{D}_i\beta^l)\tilde{\sigma}_{lj}+(\tilde{D}_j\beta^l)
\tilde{\sigma}_{li}+\beta^l(\tilde{D}_l\tilde{\sigma}_{ij})-\frac{8 \pi G }{c^4a^2}\alpha
\left(S_{ij}-\frac{1}{3}\gamma_{ij}S^l{}_l\right)+O(h\psi,h^2), \label{eqn:aw}
\end{eqnarray}
where
\begin{eqnarray}
&&\tilde{R}^{\psi}_{ij}=\frac{1}{1-2\psi}(\psi ,_{ij}+\delta_{ij}\Delta_f \psi)
+\frac{1}{(1-2\psi)^2}(3\psi,_i\psi,_j+\delta_{ij}\psi,_k\psi,_k), \label{eqn:ax}\\
&&\tilde{R}^{\psi}=\frac{1}{1-2\psi}\delta_{ij} \tilde{R}^{\psi}_{ij}\label{eqn:ay},
\end{eqnarray}
and $\Delta _f$ is the Laplacian with respect to $\delta_{ij}$.

Finally, we give the equations for matter. Since we shall restrict ourselves to  
the evolution of the density fluctuation after the decoupling of matter
and radiation, we shall adopt   
pressure-free dust as a model of the matter,
which will be a good approximation to the present universe.
The energy momentum tensor for the dust is written as 
\begin{equation}
T^{\mu\nu}=\rho c^2 u^{\mu}u^{\nu},
\end{equation}
where $u^{\mu}$ and $\rho$ are the four velocity and the density,
respectively. The $\rho$ obeys the continuity equation
\begin{equation}
(\rho u^{\mu})_{;\mu}=0.
\end{equation}
Also the 
relation between $u^i$ and $u^0$ becomes 
\begin{equation}
v^i=\frac{dx^i}{dt}=c\frac{dx^i/d\tau_p}{cdt/d\tau_p}=c\frac{u^i}{u^0},
\label{eqn:d}
\end{equation}
where $\tau_p$ represents the proper time of a dust element.
We should note that $v^i$ represents a peculiar velocity of a fluid element in
comoving frame, so the physical peculiar velocity is  $av^i$.  
 Using the peculiar velocity $v^i$, the explicit form of the continuity 
equation becomes 
\begin{equation}
\frac{1}{c}\frac{\partial \rho _{\ast}}{\partial t}+\frac{1}{c}
\frac{\partial (\rho_{\ast}v^i)}{\partial x^i}=0,
\label{eqn:b}
\end{equation}
where $\rho_{\ast}=\alpha \tilde{\gamma}^{\frac{1}{2}}a^3 \rho u^0$ is the so-called
conserved density \cite{ch,will}.
 The equations of motion are derived from 
\begin{equation} 
T^{\mu}{}_{i;\mu}=0.
\end{equation}
Making use of Eq.(\ref{eqn:b}), it takes the following form which will be 
convenient for our present purpose:
\begin{equation}
\frac{1}{c}\frac{\partial u_i}{\partial t}+\frac{v^j}{c}\frac{\partial u_i}{\partial x^j}
=-\alpha \alpha,_i u^0+\beta
^j,_iu_j-\frac{1}{2a^2}\tilde{\gamma}^{jk}{}_{,i}\frac{u_ju_k}{u^0}.
\label{eqn:az}
\end{equation}

\subsection{Cosmological post-Newtonian approximation}
\label{cosmopn}

There have been many studies 
of the post-Newtonian approximation in the cosmological
circumstances in the Eulerian
framework \cite{nu,i,peebles,tomita88,futamase88,futamase89,tomita91}.
Here we shall develop the PN approximation
based on the (3+1) formalism described above
\cite{futamase96,sa}.

We remark that the metric required to the usual Eulerian Newtonian
picture in the perturbed FRW universe 
is known to take the following form:
\begin{equation}
ds^2=-\left(1+\frac{2 \phi_N}{c^2}\right)c^2dt^2+a^2(t)\left(1-\frac{2 \phi_N}{c^2}\right)
\delta_{ij}dx^idx^j, 
\label{eqn:c}
\end{equation}
where $\phi_N $ is the Newtonian gravitational potential 
related to the matter density 
fluctuation field $\delta_N$ via the Poisson's equation:
\begin{equation}
\Delta_f \phi_N({\mbox{\boldmath $x$}},t)=4\pi G a^2(t)\rho_b(t)\delta_N(
{\mbox{\boldmath$x$}},t).
\label{eqn:ba}
\end{equation}
where $\rho_b=\kaco{\rho(\bm{x},t)}_{V_H}$ and 
\[\delta_N({\mbox{\boldmath $x$}},t)\equiv \frac{\rho_N({\mbox{\boldmath $x$}},t)-\rho_b(t)}
{\rho_b(t)}. \]
The above metric is usually used to give an accurate description of the trajectories
of nonrelativistic fluid elements such as CDM inside scales
much smaller than the Hubble distance $cH^{-1}$ (Newtonian limit). 

We shall expand the basic equations by using $c^{-1}$ as the perturbation 
parameter in order 
to have the post-Newtonian approximation under the condition that 
the lowest order of metric takes the above Newtonian form 
(\ref{eqn:c}).  Thus we adopt the coordinate conditions
where the lowest order of $\alpha$ and $\psi$ are of the order $c^{-2}$, 
 and other metric perturbations are of the order $c^{-3}$ at most. 
Therefore, according to Chandrasekhar's description \shortcite{ch}, 
the PN terms of all metric variables used in this paper may be 
expanded as follows: 
\begin{eqnarray}
&&\alpha = 1+\frac{\phi}{c^2}+\frac{\alpha^{(4)}}{c^4}+O(c^{-6}), \nonumber \\
&&\psi= \frac{\psi^{(2)}}{c^2}+\frac{\psi^{(4)}}{c^4}+O(c^{-6}),  \nonumber \\
&&\beta^i=\frac{\beta^{(3)i}}{c^3}+O(c^{-5}), \nonumber \\
&& h_{ij}=\frac{h^{(4)}_{ij}}{c^4}+O(c^{-5}), \nonumber \\
&& \tilde{\sigma}_{ij}=\frac{\tilde{\sigma}^{(3)}_{ij}}{c^3}+O(c^{-5}), \nonumber \\
&& \theta=\frac{\theta^{(3)}}{c^3}+O(c^{-5}),\label{eqn:bb}
\end{eqnarray}
where $\phi$ is the Newtonian gravitational potential as shown below. 
Note that $\psi^{(2)}$ is not same as  $\phi$ a priori, 
but actually we will see that they coincide. 
Thus the metric up to the order  $c^{-2}$ 
agrees with (\ref{eqn:c}). Also if we assume that
the lowest order in $\beta ^i$ is of the order  $c^{-3}$, 
the lowest order in $\tilde{\sigma}_{ij} $ and $\theta$ both 
become $c^{-3}$ through  Einstein equations. 
Using the above variables, the four velocity is expanded as 
\begin{eqnarray}
&&u^0=1+\left\{\frac{1}{c^2}\left( \frac{1}{2}a^2 v^2-\phi\right)\right\}
 +O(c^{-4}), \nonumber \\
&&u_0=-\left[1+\left\{\frac{1}{c^2}\left(\frac{1}{2}a^2v^2+\phi\right)\right\}+O(c^{-4})
\right], \nonumber \\
&&u^i=\frac{v^i}{c}\left[1+\left\{\frac{1}{c^2}\left(\frac{1}{2}a^2v^2-\phi\right)\right\}
 \right]
+O(c^{-5}), \nonumber \\
&&u_i=a^2\left[ \frac{v^i}{c}+\left\{\frac{v^i}{c^3}\left(
 \frac{1}{2}a^2v^2 -\phi -2\psi^{(2)}
\right)+\frac{1}{c^3}\beta^{(3)i}\right\}\right]+ O(c^{-5}),\label{eqn:bc}
\end{eqnarray}
where terms in the bracket $\{\}$ denote the first PN corrections, 
and $v^i$ is equal to that defined in Eq.(\ref{eqn:d}) and  $v^2=\delta_{ij}v^iv^j$. 
Following Shibata and Asada \shortcite{sa},  we consider only the PN expansion 
of $\beta^i$, not of $\beta_i$, since only $\beta^i$ appears 
in the above expressions. $E, J_i, S_{ij}$ and $S^l{}_l$ are likewise 
expanded as       
\begin{eqnarray}
&&E=\rho c^2\left[1+\frac{1}{c^2}a^2v^2+O(c^{-4})\right] 
,\nonumber \\
&&J_i=\rho c^2 a^2\left[ \frac{v^i}{c}+\frac{v^i}{c^3}\left( a^2v^2 -\phi -2\psi^{(2)}
\right)+\frac{1}{c^3}\beta^{(3)i}+O(c^{-5})\right] \nonumber \\
&&S_{ij}=\rho c^2 a^4\left[\frac{1}{c^2}v^iv^j+O(c^{-4})\right], \nonumber \\
&&S^i{}_i=\rho c^2 a^2 \left[ \frac{1}{c^2}v^2+O(c^{-4})\right]. \label{eqn:bd}
\end{eqnarray}

By  substituting the above expressions into Einstein equations, 
we can derive the relations between the metric variables and
the matter variables in each order of  $c^{-n}$.  
From Eqs.(\ref{eqn:au}) and (\ref{eqn:bd}), we find 
\begin{eqnarray}
&&\frac{1}{a^2}\left[ \frac{1}{c^2}\Delta_f \phi 
+\frac{1}{c^4}\left( \Delta_f \alpha^{(4)}+
2\psi^{(2)}\Delta_f\phi -\psi^{(2)},_k\phi,_k\right)\right] =
\frac{1}{c^2} 4 \pi G (\rho-\rho_b) \nonumber \\
&&\hspace{10em}+\frac{1}{c^4}\left[ 4\pi G (\rho \phi +2 \rho_b\phi+2\rho a^2
v^2  )  -\frac{\partial }{\partial t}\theta ^{(3)}-2\frac{\dot{a}}{a}\theta ^{(3)}\right]
+O(c^{-6}).
\label{eqn:be}
\end{eqnarray}
The equation for $\phi$ is derived in the $c^{-2}$ order part of the
above equation: 
\begin{eqnarray}
&&\Delta_f\phi=4 \pi G a^2(\rho -\rho_b). \label{eqn:bf}
\end{eqnarray}
Thus we may call $\phi $ the Newtonian-like 
 gravitational potential. 
Similarly from Eqs.(\ref{eqn:av}) and (\ref{eqn:bd}), 
we find the equation for $\psi^{(2)}$ as follows:
\begin{eqnarray}
&&\Delta_f \psi^{(2)}=4 \pi G a^2 (\rho-\rho_b) \label{eqn:bg}.
\end{eqnarray}
So we can conclude 
\begin{eqnarray}
&&\psi^{(2)}=\phi. \label{eqn:bh}
\end{eqnarray}
The PN potentials $\alpha^{(4)}$ and $\psi^{(4)}$ are obtained by taking the
$c^{-4}$ part of Eqs.(\ref{eqn:be}) and (\ref{eqn:av}), respectively,
\begin{eqnarray}
&&\Delta_f\alpha^{(4)}=\phi,_k\phi,_k+4\pi G a^2 \left( 2\rho a^2 v^2 -\rho \phi +4 \rho_b 
\phi \right) -a^2 \frac{\partial \theta}{\partial t}^{(3)}-2 a \dot{a}\theta^{(3)},
\label{eqn:bk}\\
&&\Delta_f\psi^{(4)}=4 \pi G a^2\left\{\rho a^2v^2-4(\rho-\rho_b)\phi\right\}
 -\frac{3}{2}\phi,_i\phi,_i+a\dot{a}\theta^{(3)},
\label{eqn:bn}
\end{eqnarray}
where we used the Newtonian order equations (\ref{eqn:bf}) and (\ref{eqn:bh}).

The relation between $\beta^{(3)i} $ and the matter variables is obtained 
from  Eqs.(\ref{eqn:ao}) and (\ref{eqn:bd}) to be 
\begin{eqnarray}
&&\Delta_f\beta^{(3)i}-\beta^{(3)j},_{ji}-4\left(\frac{ \partial \phi,_i}{\partial t}+
\frac {\dot{a}}{a}\phi,_i\right)=16 \pi G a^2 \rho v^i,
\label{eqn:bm}
\end{eqnarray}
where we used the following relation derived from  
 (\ref{eqn:aj}):
\begin{eqnarray}
&&\theta^{(3)}=3\left( \frac{\partial \phi}{\partial t}+\frac{\dot{a}}{a}\phi\right)+
\beta^{(3)i},_i, \nonumber \\
&& \tilde{\sigma}_{ij}^{(3)}=\frac{1}{2}\left(\beta^{(3)i}{}_{,j}+
\beta^{(3)j}{}_{,i}\right)-\frac{1}{3}\delta_{ij}\beta^{(3)k},_k. \label{eqn:bj}
\end{eqnarray}

Finally, we give the material equations up to the first PN order.
 From Eqs.(\ref{eqn:b}) and (\ref{eqn:az}), the continuity equation
and equations of motion become respectively 
\begin{eqnarray}
&&\frac{\partial }{\partial t}\left[ \rho a^3 \left\{ 1+\frac{1}{c^2}\left( 
\frac{1}{2}a^2v^2
-3\phi \right) \right\} \right]\!+\!\frac{\partial }{\partial x^i}\!\left[
\rho a^3 v^i \left\{ 1+\frac{1}{c^2}\left(\frac{1}{2}a^2v^2-3\phi\right) \right\}\right]
\!+O(c^{-4})\! =0, \label{eqn:bt} \\
&&\frac{\partial }{\partial t}\left[ a^2v^i\left\{1+ \frac{1}{c^2}\left(
\frac{1}{2}a^2v^2-3\phi\right)\right\}\right]+v^j\frac{\partial}{\partial x^j}
\left[ a^2v^i\left\{1+ \frac{1}{c^2}\left(\frac{1}{2}a^2v^2-3\phi\right)\right\}\right]
\nonumber \\
&&\hspace{6em}=\!-\phi,_i\!-\frac{1}{c^2}\left[\alpha^{(4)},_i+\frac{3}{2}a^2v^2\phi,_i
-a^2\beta^{(3)j},_iv^j+\frac{\partial }{\partial t}(a^2 \beta^{(3)i})+
v^j\frac{\partial}{\partial x^j}(a^2 \beta^{(3)i})\right]\!+O(c^{-4}).  \label{eqn:bp}
\end{eqnarray}
The lowest order in these two equations reduces to
the Newtonian Eulerian equations of 
hydrodynamics on the FRW background 
for a pressureless fluid \cite{peebles}, and terms of the order $c^{-2}$ provide 
the first PN corrections. 
For convenience,  Eq.(\ref{eqn:bp}) can be rewritten as 
\begin{eqnarray}
&&\frac{\partial v^i}{\partial t}+v^j\frac{\partial v^i}{\partial x^j}+2\frac{\dot{a}}{a}v^i
+\frac{1}{c^2}v^i\left[ \frac{\partial }{\partial t}\left(\frac{1}{2}a^2v^2-3\phi 
\right)+v^j\frac{\partial}{\partial x^j}\left(\frac{1}{2}a^2v^2
-3\phi\right)
\right] \nonumber \\
&&\hspace{8em}=-\frac{1}{a^2}\phi,_i-\frac{1}{c^2}\frac{1}{a^2}\left[ \alpha^{(4)},_i+
3\phi \phi,_i+a^2 v^2\phi,_i
+\frac{\partial }{\partial t}(a^2 \beta^{(3)i})+a^2v^j(\beta^{(3)i},_j-\beta^{(3)j},_i)
 \right] +O(c^{-4}).  \label{eqn:g}
\end{eqnarray}
Eqs.(\ref{eqn:bt}) and (\ref{eqn:g}) are our basic equations for
the relativistic 
Lagrangian approach to the trajectory field of  matter fluid elements.

We note that the above equations do not contain the $\Lambda$ term
explicitly, except in the equation to determine the expansion law, (\ref{eqn:at}).
Therefore, they may be used for the case $\Lambda\neq 0$ 
as well. 

Before going into the Lagrangian perturbation theory, 
we consider linearized  theory in the Eulerian picture  for
PN approximation on the Einstein-de Sitter background (${\cal
K}=\Lambda=0$).
In doing so we remark the order of linear terms of $\phi/c^2 $ and
$\alpha^{(4)}/c^4$,
respectively:  
\begin{eqnarray}
&&\frac{\phi}{c^2}\sim \frac{G \rho_b (al)^2}{c^2}\delta \sim \left(\frac{al}{ct}\right)^2
\delta ,\nonumber \\
&&\frac{\alpha^{(4)}}{c^4}\sim \frac{G \rho_b (al)^2}{c^2}\frac{\phi}{c^2} \sim \left(
\frac{al}{ct}\right)^4 \delta ,\label{eqn:zo}
\end{eqnarray}
where $l$ denotes the characteristic length of a density fluctuation 
in the comoving frame and we have used Eqs.(\ref{eqn:bf}) and (\ref{eqn:bk}).
This equation (\ref{eqn:zo}) means that  $\alpha^{(4)}$
may become larger than the quadratic terms of perturbation quantities
such as
$\delta^2$ for fluctuations with scales equivalent to 
 or beyond  the horizon scale $ct$.
Therefore, we will leave all linear terms of perturbation 
quantities in the above equations regardless of
the power of $c$.

By linearizing  Eqs.(\ref{eqn:bt}) and (\ref{eqn:g}), we get the following equations:
\begin{eqnarray}
&& \frac{\partial}{\partial t}\left(\delta-\frac{3\phi}{c^2}\right)
+v^i{}_{,i}=0, \label{eqn:zp}\\
&&\frac{\partial}{\partial t}v^i+2\frac{\dot{a}}{a}v^i=-\frac{1}{a^2}\phi_{,i}
-\frac{1}{c^2}\frac{1}{a^2}\left[\alpha^{(4)}{}_{,i}+\frac{\partial }{\partial t}(
a^2\beta^{(3)i})\right] \label{eqn:zq},
\end{eqnarray} 
where we used the fact that $\rho_b(t)a(t)^3$ is constant. Using  Eqs.(\ref{eqn:bf}),
(\ref{eqn:bk})
and (\ref{eqn:bj}), Eq.(\ref{eqn:zq}) becomes
\begin{eqnarray}
&&\frac{\partial}{\partial t}v^i{}_{,i}+2\frac{\dot{a}}{a}v^i{}_{,i}=-4\pi G \rho_b
 \delta -\frac{1}{c^2}4\pi G \rho_b 2\phi +\frac{3}{c^2}\left(
 \frac{\partial^2\phi}{\partial t^2}+3\frac{\dot{a}}{a}\frac{\partial \phi}{\partial t}
 \right).\label{eqn:zr}
\end{eqnarray}
By substituting Eq.(\ref{eqn:zp}) into Eq.(\ref{eqn:zr}), we can get
the evolution equation for the density contrast in our coordinate system:
\begin{eqnarray}
&&\frac{\partial ^2 \delta}{\partial t^2}+2\frac{\dot{a}}{a}\frac{\partial \delta}{
\partial t}-4\pi G \rho_b \delta=\frac{1}{c^2}4\pi G\rho_b\cdot2\phi+\frac{3}{c^2}
\frac{\dot{a}}{a}\frac{\partial \phi}{\partial t}. \label{eqn:zs}
\end{eqnarray}
As we will show in the next section, $\phi $ remains constant during a period
when  the linear approximation is valid. Thus 
we can integrate Eq.(\ref{eqn:zs}) to get
\begin{eqnarray}
&&\delta({\mbox{\boldmath$x$}},t)=\left(\frac{t}{t_I}\right)^{2/3}\left(\delta(
{\mbox{\boldmath$x$}},t_I)+\frac{2}{c^2}\phi({\mbox{\boldmath$x$}})\right)
-\frac{2}{c^2}\phi(\mbox{\boldmath$x$}) \label{eqn:zt},
\end{eqnarray}   
where $t_I$ denotes the initial time and
we have considered only the growing mode because 
the decaying mode ($\propto t^{-1}$)
plays physically no important role.
It is remarked that $(\delta+2\phi/c^2)$ is a gauge invariant quantity in
the PN approximation and is guaranteed to be a physical density
fluctuation (see Appendix A). 
In next section we will use Eq.(\ref{eqn:zt}) as the bridge connecting
between the Eulerian perturbation  picture and the  Lagrangian 
perturbation picture
in order to determine
the initial conditions of the first PN displacement vector.

\section{Lagrangian perturbation approach to the perturbed FRW universe 
in the first post-Newtonian approximation}

\subsection{Post-Newtonian Lagrangian evolution equations for the trajectory field}
\label{rellagevo}

In this section our purpose is to construct the complete set of 
evolution equations for the trajectory field of the 
the fluid elements  in the post-Newtonian 
approximation by extending  the formalism in Newtonian theory
developed by
Buchert.
 
In the following discussion, we only consider gravitational 
instability on
the Einstein-de Sitter background (${\cal K}=\Lambda=0$) for simplicity:
\begin{equation}
H^2(t)=\frac{8}{3}\pi G \rho_b(t). \label{eqn:zm}
\end{equation}
The generalization to more general FRW background is straightforward 
and will 
be presented elsewhere. 

In the Lagrangian description, we concentrate on the integral 
curves $\bm{x}=\bm{f}(\bm{X},t)$
of the velocity
 field $\bm{v}(\bm{x},t)$:
\begin{equation}
\frac{d \bm{f}}{dt} \equiv \left.\frac{\partial \bm{f}}
{\partial t}\right|_X
=\bm{v}(\bm{f},t) ,\hspace{1.5cm} \bm{f}(\bm{X},t_I)\equiv \bm{X},
\label{eqn:map}
\end{equation}
where $\bm{X}$ denote the Lagrangian coordinates 
which label fluid elements, \bm{x} are 
the positions of these elements in Eulerian space at time $t$, 
and $t_I$ is the initial time when  Lagrangian coordinates are defined. 
It should be noted that we introduce the Lagrangian coordinates on the
comoving coordinates because we have already derived basic equations by
using  FRW metric defined with 
the comoving coordinates. 
As long as the mapping $\bm{f}$ 
is invertible,
 we can give the inverse of the deformation tensor $f_{i|j}$ which
is written in terms of  variables $(\bm{X},t)$:
\begin{equation}
\left.\frac{\partial X^i}{\partial x^j}\right|_t \equiv h_{i,j}(\bm{X},t) 
=\frac {1}{2J}\epsilon_{iab}\epsilon_{jcd}f_{c|a}f_{d|b} ,\label{eqn:dm}
\end{equation}
where $J$ is the determinant of the deformation tensor  $f_{i|j}$,
and  $\epsilon_{ijk}$ is the totally antisymmetric quantity with
$\epsilon_{123}=+1$. 
The comma and the vertical slash  in the subscript denote 
partial differentiation with respect 
to the Eulerian coordinates and the Lagrangian coordinates, 
respectively. 
The following equations are naturally satisfied:
\begin{equation} 
h_{i,k}f_{k|j}=\delta_{ij},\hspace{2em}f_{i|k}h_{k,j}=\delta_{ij}, 
\label{eqn:definv}
\end{equation}
where $\delta_{ij}$ denotes Kronecker's delta function.
Using the above equations and the following identities
\begin{eqnarray}
&&\frac{d}{dt}\equiv \left.\frac{\partial }
{\partial t}\right|_X= \left.\frac{\partial}{\partial t}\right|_x+v^i
\left.\frac{\partial }{\partial x^i}\right|_t, \nonumber \\
&&\frac{1}{J(\bm{X},t)}\frac{d}{dt}J(\bm{X},t)=\left.
\frac{\partial X^k}{\partial x^i}\right|_t\left.\frac{\partial v^i(\bm{X},t)}
{\partial X^k}\right|_t, 
\end{eqnarray}
the continuity equation (\ref{eqn:bt}) in the PN approximation
may be written as 
\begin{equation}
\frac{d}{dt}\left[ \rho a^3\left(1+\frac{1}{c^2} A(\bm{X},t)
\right)J(\bm{X},t)\right]
+O(c^{-4})=0 ,\label{eqn:da}
\end{equation}
where 
\begin{equation}
A(\bm{X},t):=\frac{1}{2}a^2(t)\left(\frac{d\bm{f}}{dt}\right)^2
-3\phi(\bm{X},t).
\end{equation}
Thus the density field is integrated exactly up to the first PN order 
in the Lagrangian picture as in the Newtonian case:
\begin{equation}
\rho(\bm{X},t)a^3J(\bm{X},t)=\left(1+ \frac{1}{c^2}A(\bm{X},t)\right)^{-1}
\mal{\rho}(\bm{X})\mal{a}^3\left(1+\frac{1}{c^2}\mal{A}(\bm{X})\right)
+O(c^{-4}),
\label{eqn:db}
\end{equation} 
where the quantities with $\mal{}$ such as $\mal{a}$ are the quantities 
at the initial time $t_I$, and 
\begin{eqnarray}
&& \mal{A}(\bm{X})=\frac{1}{2}\mal{a}^2\mal{v}^2(\bm{X})-3\mal{\phi}
(\bm{X}), \nonumber \\
&&\Delta_X\mal{\phi}(\bm{X})=4 \pi G \mal{a}^2(\mal{\rho}
(\bm{X})-\mal{\rho_b}),\nonumber
\end{eqnarray}
with $\Delta_X$ being the Laplacian with respect to the Lagrangian 
coordinates.
We should remark that the $\rho$ in all  equations below will 
appear in terms of 
$J\rho$.

Next we consider the trajectory field \bm{f} of the pressureless
fluid elements in PN approximation. The field  \bm{f} is given  by
solving Eq.(\ref{eqn:g}). The equations
 of motion 
 can be rewritten formally as 
\begin{eqnarray}
&&\frac{d^2f^i}{dt^2}+2\frac{\dot{a}}{a}\frac{df^i}{dt}={}^{(N)}\!g^i(\bm{X},t)
+\frac{1}{c^2}{}^{(PN)}\!g^i(\bm{X},t)+\frac{1}{c^4}{}^{(2PN)}\!g^i
(\bm{X},t)+\cdots,\label{eqn:dc}
\end{eqnarray} 
where ${}^{(N)}\!g^i$,  ${}^{(PN)}\!g^i$ and  ${}^{(2PN)}\!g^i$ denote the 
acceleration fields in
the order $c^0$ ,$c^{-2}$, and the  order $c^{-4}$, respectively. 
By means of the perturbation theory,
the trajectory field given by integrating the above equation may
be written  formally as
\begin{eqnarray}
&&\bm{f}(\bm{X},t)=\bm{X}+{}^{(N)}\!\bm{p}(\bm{X},t)+\frac{1}{c^2}
{}^{(PN)}\!\bm{p}(\bm{X},t)+\frac{1}{c^4}{}^{(2PN)}\!\bm{p}(\bm{X},t)
+\cdots, \label{eqn:dd}
\end{eqnarray}
where \bm{p} denotes the displacement vector and the following
initial conditions 
 must be imposed by Eq.(\ref{eqn:map}): 
\begin{eqnarray}
&&{}^{(N)}\!\bm{p}(\bm{X},t_I)={}^{(PN)}\!\bm{p}(\bm{X},t_I)=\cdots=\mbox{{\bf 0}}.
\end{eqnarray}

Now we construct the set of equations to solve the  
 trajectory field $\bm{f}$.  Likewise as a Newtonian case \cite{buchertpre}, 
we consider the divergence and the rotation  of equations 
of motion in PN approximation with respect to the Eulerian coordinates,
 respectively. 
In the Lagrangian picture, the former set becomes basically  
the evolution equation of the longitudinal part of the trajectory 
field and the latter becomes one of its transverse part. 
First  by operating divergence on Eq.(\ref{eqn:g}) with respect to 
the {\it Eulerian} coordinates and using Einstein equations in each order of
$c$,  we obtain
\begin{eqnarray}
\lefteqn{\frac{\partial }{\partial x^i}\left[ \frac{d^2f^i}{dt^2} 
\right] +2\frac{\dot{a}}{a}
\frac{\partial }{\partial x^i}\left[ \frac{df^i}{dt} \right]} \nonumber \\
&&\hspace{3em}=-4 \pi G \rho \left(1+\frac{1}{c^2}A\right)+4\pi G \rho_b-
\frac{1}{c^2}\frac{\partial }{\partial x^i}
\left[ \frac{df^i}{dt}\frac{dA}{dt}\right]
-\frac{1}{c^2}\left[ \frac{4}{a^2}\phi_{,i}\phi_{,i}
 +4\pi G\left\{ \rho\left(-\frac{3}{2} a^2v^2+5\phi\right) 
+\rho_b\left(- a^2v^2 + \phi \right)\right\}\right.\nonumber \\
&&\hspace{4em}\left.+(v^2)_{,i}\phi_{,i}
-\frac{3}{a^2}\frac{d}{dt}\left\{ a^2
\left( \frac{\partial \phi}{\partial t}+\frac{\dot{a}}{a}\phi\right)\right\}
-v^i \left( \frac{\partial \phi _{,i}}{\partial
      t}+\frac{\dot{a}}{a}\phi_{,i}\right)+v^j{}_{,i}\left(\beta^{(3)i}{}_{,j}
-\beta^{(3)j}{}_{,i}\right)
      \right]\!\!+\!O(c^{-4}). \label{eqn:dg} 
\end{eqnarray}
According to the usual 
procedure in the Lagrangian formalism, 
 we modify the above equation by using Eqs.(\ref{eqn:map}),
(\ref{eqn:dm}) and (\ref{eqn:definv}) in the following form: 
\begin{eqnarray}
\lefteqn{\frac{1}{2}\epsilon_{jab}\epsilon_{icd}f_{c|a}f_{d|b}
\left[ \frac{d^2f^i}{dt^2}+2\frac{\dot{a}}{a} \frac{df^i}{dt}
 \right]}  \nonumber \\
&&\hspace{1em}=-4 \pi G \frac{1}{a^3}\mal{\rho }\mal{a}^3
 \left(1+\frac{1}{c^2}\mal{A}\right)
+4\pi G \rho_bJ-\frac{1}{2c^2}\epsilon_{jab}\epsilon_{icd}f_{c|a}
f_{d|b}\frac{\partial }{\partial X^j}\left[ \frac{df^i}{dt}
\frac{dA}{dt}\right]\nonumber\\
&&\hspace{2em}-\frac{1}{c^2J}\epsilon_{jab}\epsilon_{kpq}f_{c|a}f_{d|b}
f_{c|p}f_{d|q}\left\{\frac{2}{a^2}\phi_{|j}\phi_{|k}+\frac{df_{l|j}}
{dt}\frac{df_l}{dt}
\phi_{|k}\right\}\!-\!\frac{4\pi G }{c^2}\!
\left[ J\rho\left\{-\frac{3}{2} a^2\left(\frac{df_i}{dt}\right)^2\!
+5\phi \right\}
+J\rho_b\left\{- a^2\left(\frac{df_i}{dt}\right)^2
+ \phi \right\}\right]\!\nonumber \\
&&\hspace{2em} +\!\frac{3J}{a^2c^2}\frac{d}{dt}\left\{ a^2
\left( \frac{d \phi}{d t}-\frac{1}{2J}\epsilon_{jab}\epsilon_{icd}
f_{c|a}f_{d|b}\phi_{|j}\frac{df_i}{dt}+\frac{\dot{a}}{a}
\phi\right)\!\right\}\nonumber\\
&&\hspace{2em}+\frac{J}{c^2}\frac{df_i}{dt}\left(
\frac{d}{dt}-\frac{1}{2J}
\epsilon_{lab}\epsilon_{kcd}f_{c|a}f_{d|b}\frac{df_k}{dt}
\frac{\partial}{\partial X^l}+\frac{\dot{a}}{a}
\right)\frac{1}{2J}\epsilon_{mpq}\epsilon_{irs}f_{r|p}f_{s|q}\phi_{|m}
-\frac{J}{c^2}v^j{}_{,i}\left(\beta^{(3)i}{}_{,j}
-\beta^{(3)j}
{}_{,i}\right)+\!O(c^{-4}), \label{eqn:dh}
\end{eqnarray}
where 
\[
J\equiv {\rm det }\left(\bm{X}+{}^{(N)}\!
\bm{p}+\frac{1}{c^2}{}^{(PN)}\!\bm{p}+\cdots \right),
\]
and  we have  used Eq.(\ref{eqn:db}), which was obtained by solving the
continuity equation, and the Eulerian partial time derivative of
$\phi$ was rewritten as, 
for example, 
\begin{eqnarray}
\frac{\partial \phi}{\partial t}&=&\frac{d \phi}{dt}-
v^i\frac{\partial \phi}{\partial x^i}\nonumber \\
&=&\frac{d \phi}{dt}
-\frac{df_i}{dt}\frac{1}{2J}\epsilon_{jab}\epsilon_{icd}f_{c|a}
f_{d|b}\phi_{|j}
.\nonumber 
\end{eqnarray}
We did not transform  the terms including the shift vector
$\beta^{(3)i}$  in Eq.(\ref{eqn:dh}) 
into the form in Lagrangian picture  on purpose,
because its expression in Eulerian picture 
will make us  clearer  to understand the expansion of 
the above equation introduced 
in later discussions. 

Next by considering the  rotation of Eq.(\ref{eqn:g}) with respect to the 
Eulerian coordinates, we get the following equation: 
\begin{eqnarray}
&&\hspace{-2em}\epsilon_{ijk}\frac{\partial }{\partial x^j}\!\left[
\frac{d^2 f^k}{dt^2}+2\frac{\dot{a}}{a}\frac{d f^k}{dt}+\frac{1}{c^2}
\frac{df^k}{dt}\frac{d A}{dt}\right] \nonumber \\
&&\hspace{0em}=-\frac{1}{c^2}\frac{1}{a^2}
\epsilon_{ijk}\frac{\partial }{\partial x^j}\left[ 
a^2v^2\phi_{,k}+\frac{\partial }{\partial t}\!\left(a^2\beta^{(3)k}
\right)+a^2v^l\left(\beta^{(3)k}{}_{,l}-\beta^{(3)l}{}_{,k}\right)
\right]+O\!\left(c^{-4}\right),
\end{eqnarray}
Similarly  
we again modify the above equation by using
Eqs.(\ref{eqn:map}), (\ref{eqn:dm}) and (\ref{eqn:definv}) in the following form:  
\begin{eqnarray}
\lefteqn{\epsilon_{abc}f_{j|a}f_{i|b}\left[\frac{d^2 f_{j|c}}{dt^2}+
2\frac{\dot{a}}{a}\frac{df_{j|c}}{dt}\right]}\nonumber \\
&&=
-\frac{J}{c^2}\left[\frac{1}{a^2}\frac{\partial }{\partial t}
\!\left(a^2\epsilon_{ijk}\beta^{(3)k}{}_{,j}\right)
+\epsilon_{ijk}\frac{\partial }{\partial x^j}\!
\left\{v^l\left(\beta^{(3)k}{}_{,l}-\beta^{(3)l}{}_{,k}\right)\right\}
\right.\nonumber \\
&&\hspace{1em}\left.+\frac{2}{J}\epsilon_{abc}f_{i|b}\frac{df_l}{dt}
\frac{df_{l|c}}{dt}\phi_{|a}\right]-\frac{1}{c^2}\epsilon_{abc}f_{j|a}
f_{i|b}\left(\frac{df_{j|c}}{dt}\frac{d A}{dt}+\frac{df_j}{dt}
\frac{dA_{|c}}{dt}\right)+O\!\left(c^{-4}\right), \label{eqn:transtrajdiff}
\end{eqnarray}
where the term containing 
the Eulerian partial spatial derivative of $\phi$ was rewritten
as
 \begin{eqnarray}
&&\epsilon_{ijk}\frac{\partial }{\partial x^j}\!\left(
v^2\phi_{,k}\right)
=\frac{2}{J}\epsilon_{abc}f_{i|b}\frac{df_m}{dt}\frac{df_{m|c}}{dt}
\phi_{|a}.\nonumber 
\end{eqnarray}
We note that the right-hand side of Eq.(\ref{eqn:transtrajdiff}) 
is of the order $c^{-2}$, namely, has  only 
the first PN order quantities since we consider the gravitational 
instability of the self-gravitating {\it dust} model. 
Eqs.(\ref{eqn:dh}) and (\ref{eqn:transtrajdiff})
 are our basic equations 
for the trajectory field $f_i$ in the Lagrangian picture 
in the PN approximation. 
These are evolution equations for general trajectory fields which can be
applied even in the nonlinear situation ($\delta\gg 1$) 
as long as the PN approximation introduced in \S \ref{cosmopn} is valid.  
The structure of these equations allows us to solve 
the displacement vector $\bm{p}$  iteratively up to any desired 
order in $c^{-n}$ in terms of the lower order displacement vectors and metric variables.  
(The shift vector is expressed by solving Eqs.(\ref{eqn:bm}) and (\ref{eqn:bj})
in terms of Newtonian order quantities such as $\phi$.)
Thus we need to know only the solution at the Newtonian order   
to have the solution in the first PN approximation. 

\subsection{The irrotational flow for the trajectory field in PN approximation}
In this section, we investigate  the irrotational flow for
the trajectory field.

\subsubsection{The basic equation}
When we consider 
 irrotational flow up to PN order,
we need only  Eq.(\ref{eqn:dh}) among the above basic equations. 
Further simplification is available in this situation. 
In Newtonian case,  only an  irrotational solution
survives 
among general solutions by employing  the appropriate 
initial conditions that the
initial peculiar velocity
field is proportional to the initial peculiar gravitational field on
basis of the results derived by Newtonian  linearized
theory \cite{buchert93,buchert94}.
Thus we consider the following trajectory field with the longitudinal
perturbations:
\begin{equation}
f^i(\bm{X},t)\equiv X^i+^{(N)}\!S_{|i}(\bm{X},t)
+\frac{1}{c^2}{}^{(PN)}\!S_{|i}(\bm{X},t)+\cdots. \label{eqn:zc}
\end{equation}
Note that the following initial conditions must be imposed:
\[ {}^{(N)}\!S_{|i}(\bm{X},t_I)={}^{(PN)}\!S_{|i}(\bm{X},t_I)
=\cdots=0.\]
It should be noted that the above expansion of the trajectory field
 here is a post-Newtonian expansion introduced  through
the equations of motion in PN approximation as discussed in the previous 
section, and 
we impose no restriction on the amplitude of density contrast field. 
Neglecting the vorticity 
is known to be a reasonable assumption for the 
pressureless fluid in Newtonian  theory,
but this is not the case for the PN displacement vector. 
We shall discuss the PN transverse flow  in our coordinates later  and 
here concentrate only on the irrotational flow.
Inserting  Eq.(\ref{eqn:zc}) into Eq.(\ref{eqn:dh}), we obtain the following
equation  from  the $c^0$  order:
\begin{equation}
\frac{1}{2}\epsilon_{jab}\epsilon_{icd}{}^{(N)}\!f_{c|a}{}^{(N)}\!f_{d|b}
\left[{}^{(N)}\!\ddot{f}_{i|j}+2\frac{\dot{a}}{a}{}^{(N)}\!\dot{f}_{i|j}
\right]
=-\frac{4\pi G}{a^3}\mal{a}^3\left(\mal{\rho}-\mal{\rho}_bJ_N\right),
\label{eqn:ln}
\end{equation}
where 
\begin{eqnarray}
&& {}^{(N)}\!f_i(\bm{X},t)\equiv X^i+{}^{(N)}\!S_{|i}(\bm{X},t),
\nonumber \\
&& J_N\equiv {\rm det}\left({}^{(N)}\!f_{i|j}\right)={\rm det}\left(
\delta_{ij}+{}^{(N)}\!S_{|ij}\right),\nonumber 
\end{eqnarray}
and the dot denote the Lagrangian  time-derivative.
In this case,  Eq.(\ref{eqn:ln}) entirely agrees with the result derived 
from the following familiar set of  equations  in  
Newtonian  theory in the Lagrangian description,
\begin{eqnarray}
&&\rho_Na^3=\frac{\mal{\rho}\mal{a}^3}{J_N}, \label{eqn:zh} \\
&&\frac{d}{dt}\left[ a^2\frac{d{}^{(N)}\!f_{i}}{dt} \right]=
-\frac{1}{2J_N}\epsilon_{jab}\epsilon_{icd}{}^{(N)}\!f_{c|a}
{}^{(N)}\!f_{d|b}\phi_{N|j},
 \label{eqn:df}\\
&&\frac{1}{2}\epsilon_{jab}\epsilon_{kpq}
{}^{(N)}\!f_{c|a}{}^{(N)}\!f_{d|b}
\frac{\partial}{\partial X^j}\left(\frac{1}{J_N}{}^{(N)}\!
f_{c|p}{}^{(N)}\!f_{d|q}
\phi_{N|k}\right)=\frac{4\pi G}{a}  \left( \mal{\rho}
\mal{a}^3-J_N\rho_b 
a^3\right),
\end{eqnarray}
that is, these equations can be changed to the following system of
equations in the Eulerian description:
\begin{eqnarray}
&&\frac{\partial( \rho_Na^3)}{\partial t} +\frac{\partial
 (\rho_Na^3  v^i_N)}
{\partial x^i}=0, \\
&& \frac{\partial  v_N^i }{\partial t}+ v^j_N\frac{\partial v^i_N}
 {\partial x^j}+2\frac{\dot{a}}{a}v^i_N=-\frac{1}{a^2}\phi_{N,i},\\
&&\Delta_x\phi_N=4\pi G a^2\left(\rho_N-\rho_b\right) 
\end{eqnarray}
where  it is noted that a physical peculiar velocity
of a fluid element is $a(t)v^i_N$.

Now, we derive the equation which governs the first PN irrotational
velocity potential
${}^{(PN)}\!S$. For this purpose we must express terms 
in the right-hand side of Eq.(\ref{eqn:dh}) explicitly by Newtonian 
quantities. This will be easily done by using Eq.(\ref{eqn:db}) and 
assuming 
\begin{equation}
J_N > \frac{1}{c^2}\tilde{J}_{PN}, \label{eqn:deterassume}
\end{equation}
where $\tilde{J}_{PN}$ is a term of the order $c^{-2}$  in expanding $J$: 
\begin{equation}
\tilde{J}_{PN}={}^{(PN)}\!{S}_{|ii}+\left(
{}^{(PN)}\!{S}_{|ii}{}^{(N)}\!{S}_{|jj}-{}^{(PN)}\!{S}_{|ij}
{}^{(N)}\!{S}_{|ij}\right)+\frac{1}{2}\epsilon_{iab}
\epsilon_{jcd}{}^{(PN)}\!{S}_{|ca}{}^{(N)}\!{S}_{|db}
{}^{(N)}\!{S}_{|ij}. 
\end{equation}
(The assumption (\ref{eqn:deterassume}) seems to be  satisfied 
 almost everywhere in 
the situation  we are interested in here, 
but it may break down  at a shell-crossing point,
where the Lagrangian perturbation theory itself also do.
We leave this
problem for the present.)    
According to the above discussion, for example, we shall make the following replacement 
in the right hand side of Eq.(\ref{eqn:dh}):
\begin{eqnarray}
\lefteqn{\frac{1}{c^2J}\epsilon_{jab}\epsilon_{kpq}
f_{c|a}f_{d|b}f_{c|p}f_{d|q}
\frac{df_{l|j}}{dt}\frac{df_l}{dt}\phi_{|k}}\nonumber \\
&&\hspace{1em}=\frac{1}{c^2J_N}\epsilon_{jab}
\epsilon_{kpq}{}^{(N)}\!f_{c|a}{}^{(N)}\!f_{d|b}{}^{(N)}\!f_{c|p}{}^{(N)}\!
f_{d|q}\frac{d{}^{(N)}\!f_{l|j}}{dt}\frac{d{}^{(N)}\!f_l}{dt}
\phi_{N|k}+O\left(
c^{-4}\right)
\nonumber \\
&&\hspace{1em}=\frac{1}{2c^2J_N}\epsilon_{jab}\epsilon_{icd}\epsilon_{irs}
\epsilon_{kpq}{}^{(N)}\!f_{c|a}{}^{(N)}\!f_{d|b}{}^{(N)}\!f_{r|p}{}^{(N)}\!
f_{s|q}\frac{d{}^{(N)}\!f_{l|j}}{dt}\frac{d{}^{(N)}\!f_l}{dt}\phi_{N|k}
+O(c^{-4})\nonumber \\
&&\hspace{1em}=-\frac{1}{c^2}\epsilon_{jab}
\epsilon_{icd}{}^{(N)}\!f_{c|a}{}^{(N)}\!f_{d|b}
\frac{d{}^{(N)}\!f_{l|j}}{dt}\frac{d{}^{(N)}\!f_l}{dt}
\frac{d}{dt}\left[a^2\frac{d{}^{(N)}\!f_{i}}{dt}\right]
+O\left(c^{-4}\right).
\end{eqnarray}
where we have used Eq.(\ref{eqn:df}). Similar calculations lead us to 
the equation for the first PN displacement vector with the longitudinal
perturbation we are looking for:
\begin{eqnarray}
\lefteqn{\frac{1}{2}\epsilon_{jab}\epsilon_{icd}
{}^{(N)}\!f_{c|a}{}^{(N)}\!f_{d|b}\left[
{}^{(PN)}\!\ddot{S}_{|ij}
+2\frac{\dot{a}}{a}{}^{(PN)}\!\dot{S}_{|ij}\right]
+\tilde{I\!I}({}^{(N)}\!S,{}^{(PN)}S)+\tilde{ I\!\!I\!\!I }
({}^{(N)}S,{}^{(N)}\!S,{}^{(PN)}\!S)}
\nonumber \\
&&\hspace{0em}=-\frac{4\pi G}{a^3}\mal{\rho}\mal{a}^3\mal{A}+4\pi G\rho_b
\left[\tilde{J}_{PN}+a^2\left( {}^{(N)}\!\dot{S}_{|i}\right)^2J_N\right]
+\frac{4\pi G}{a^3}\mal{\rho}\mal{a}^3\left[\frac{3}{2}a^2
\left( {}^{(N)}\!\dot{S}_{|i}\right)^2-5\phi_N\right] \nonumber \\
&&\hspace{1em}+3J_N\!\left[\frac{d^2\phi_N}{dt^2}
\!+3\frac{\dot{a}}{a}\frac{d\phi_N}{dt}\right]\!+\!J_N\!
\left[-\frac{1}{a^2}\!\left( \frac{d (a^2
 {}^{(N)}\!\dot{S}_{|i})}{dt}\right)^2\!\!
 +2{}^{(N)}\!\dot{S}_{|i}
\frac{d^2(a^2{}^{(N)}\!\dot{S}_{|i})}{dt^2}-\frac{\dot{a}}{a}{}^{(N)}\!\dot{S}_{|i}
\frac{d(a^2{}^{(N)}\!\dot{S}_{|i})}{dt} \right]\nonumber \\
&&\hspace{1em}+\frac{1}{2}\epsilon_{jab}\epsilon_{icd}{}^{(N)}\!f_{c|a}{}^{(N)}\!f_{d|b}
\left[-\frac{\partial }{\partial
 X^j}\!\left({}^{(N)}\!\dot{S}_{|i}\dot{A}_N\right)
+\!\left({}^{(N)}\!
\dot{S}_{|k}\right)^2{}_{|j}\frac{d(a^2{}^{(N)}\!\dot{S}_{|i})}{dt}
\!+{}^{(N)}\!\dot{S}_{|k}{}^{(N)}\!\dot{S}_{|i}
\frac{d(a^2{}^{(N)}\!\dot{S}_{|kj})}{dt}\right]\!, \label{eqn:dj}
\end{eqnarray}
where
\begin{eqnarray}
&&\tilde{I\!I}({}^{(N)}\!{S},{}^{(PN)}\!S)={}^{(PN)}\!{S}_{|ii}
\left({}^{(N)}\!\ddot{S}_{|jj}
+2\frac{\dot{a}}{a}{}^{(N)}\!\dot{S}_{|jj}\right)-{}^{(PN)}\!{S}_{|ij}
\left({}^{(N)}\!\ddot{S}_{|ij}
+2\frac{\dot{a}}{a}{}^{(N)}\!\dot{S}_{|ij}\right), \nonumber \\
&&\tilde{I\!\!I\!\!I}({}^{(N)}\!{S},{}^{(N)}\!{S},{}^{(PN)}\!{S})=
\epsilon_{jab}\epsilon_{icd}
{}^{(PN)}\!{S}_{|ca}{}^{(N)}\!{S}_{|db}\left( {}^{(N)}\!\ddot{S}_{|ij}
+2\frac{\dot{a}}{a}{}^{(N)}\!\dot{S}_{|ij}\right).  
\end{eqnarray}
Here, as expected, all quadratic terms   
on the right-hand side are expressed only by Newtonian quantities. 
Since we have so far assumed that Newtonian flow is 
 irrotational, the $\beta^i_N$ terms in 
Eq.(\ref{eqn:dh})
are canceled out in Eq.(\ref{eqn:dj}), that is,
we can conclude that the gauge conditions do not have an
influence on the trajectory field in the first PN approximation.
On the contrary, 
the assumption (\ref{eqn:zc}) may mean that we tacitly 
impose gauge conditions that   there exist
 coordinates which  have  the hypersurface  normal to a 
congruence of geodesics of
the pressureless fluid elements.       
Eqs.(\ref{eqn:ln}) and (\ref{eqn:dj}) are
the master equations for solving the
longitudinal perturbations ${}^{(N)}\!S $ and ${}^{(PN)}\!S$, 
respectively. It
should be again noted that  
since the continuity equation is exactly solved,
these  equations can be integrated for some initial conditions
only if we give  the initial density field $\mal{\rho}$ 
as  the source  function.

\subsubsection{Longitudinal perturbation solutions up to 1st PN order}

Let us now solve the equations Eqs.(\ref{eqn:ln}) and (\ref{eqn:dj}).
We  first consider  Eq.(\ref{eqn:ln}) 
which is a  master equation for solving  ${}^{(N)}\!S$:
\begin{equation}
\frac{1}{2}\epsilon_{jab}\epsilon_{icd}{}^{(N)}\!f_{c|a}{}^
{(N)}\!f_{d|b}\left[
{}^{(N)}\!\ddot{f}_{i|j}+2\frac{\dot{a}}{a}{}^{(N)}\!\dot{f}_{i|j}\right]=
-\frac{4\pi G}{a^3}\mal{a}^3\left( \mal{\rho}
-\mal{\rho}_bJ_N\right).\label{eqn:dn}
\end{equation}
The Einstein-de Sitter background solution of Eq.(\ref{eqn:zm}) 
can be chosen as
\begin{equation}
a(t)=\left(\frac{t}{t_0}\right)^{2/3},
\end{equation}
where $t_0$ denotes the present time.
We use this normalization for the scale factor since it 
makes the physical interpretation of the solution more 
transparent partly because 
the length in the comoving frame is equal to the physical 
length at the present
universe with this choice. 
We shall assume that the initial hypersurface is sufficiently early 
so that the initial density contrast field is much smaller than unity. 
This allows us to adopt the order of the density contrast, say, 
$\lambda$ as a 
new perturbation parameter. Thus we  calculate the
following  ansatz for the longitudinal perturbations superposed 
on an isotropic homogeneous deformation:
\begin{equation}
{}^{(N)}\!f_i=X^i+{}^{(N)}S_{|i}=X^i
+\lambda q_z(t) \varphi^{(1)}_{|i}(\bm{X})+
\lambda^2q_{zz}(t)\varphi^{(2)}_{|i}(\bm{X})
+\cdots,
\label{eqn:dl}
\end{equation}
where the fist term on the right-hand side represents an isotropic 
homogeneous flow for the FRW background. 
The parameter $\lambda$ is assumed to be small and dimensionless. 
As described above, it can be 
considered as the amplitude of the initial density fluctuation field. 
We formally split the {\it initial} density field accordingly as 
\begin{equation}
\mal{\rho}\!(\bm{X})=\mal{\rho}_b+\lambda
 \mal{\rho}_b\,\mal{\delta}\!(\bm{X}),
 \hspace{2em}\kaco{\mal{\rho}\!(\bm{X})}_{\mal{V}}=\mal{\rho}_b,
\end{equation}
where $\mal{\delta}$ denotes the initial density contrast field. 
This choice of the initial data is adequate, since the density
need not be perturbed in the Lagrangian framework.
For simplicity
 we make use of the usual assumption that the initial
peculiar velocity is  proportional to the initial peculiar
acceleration \cite{buchert92,buchert94}. This assumption has been proven to be
appropriate  in  Newtonian linearized theory (see e.g., \cite{peebles}). 
The Newtonian peculiar velocity field${}^{(N)}\!\bm{u}(\bm{X},t)$
and acceleration field${}^{(N)}\!\bm{w}(\bm{X},t)$ is defined  as the
following forms in terms of the trajectory field $\bm{f}(\bm{X},t)$ on
the comoving frame:  
\begin{eqnarray}
&&{}^{(N)}\!\bm{u}(\bm{X},t)=a(t){}^{(N)}\!\dot{\bm{f}}(\bm{X},t), 
\nonumber \\
&&{}^{(N)}\!\bm{w}(\bm{X},t)=2\dot{a}(t){}^{(N)}\!\dot{\bm{f}}
(\bm{X},t)+a(t){}^{(N)}
\ddot{\bm{f}}(\bm{X},t)\left(=-\frac{1}{a}\nabla_x\phi_N\right).
\end{eqnarray}
Therefore we employ the following assumption for the initial data 
\[
 \mal{\bm{u}}(\bm{X})=t_I\mal{\bm{w}}(\bm{X}),
\]
so the initial conditions for $q_z\varphi^{(1)}$ 
and $q_{zz}\varphi^{(2)}$
become  
\begin{eqnarray}
&& q_z(t_I)\varphi^{(1)}_{|i}(\bm{X})=0, \hspace{1cm}q_{zz}(t_I)
\varphi^{(2)}_{|i}(\bm{X})=0, \nonumber \\
&& \lambda a(t_I)\dot{q}_z(t_I)\varphi^{(1)}_{|i}(\bm{X})
={}^{(N)}\!\mal{w}^i\!(\bm{X})t_I,
\hspace{1cm}\dot{q}_{zz}(t_I)\varphi^{(2)}
_{|i}(\bm{X})=0,
\nonumber \\
&& \lambda \left(2\dot{a}(t_I)\dot{q}_z(t_I)+a(t_I)
\ddot{q}_z(t_I)\right)
\varphi^{(1)}_{|i}(\bm{X})={}^{(N)}\!\mal{w}^i\!(\bm{X}), 
\end{eqnarray}
where ${}^{(N)}\!\mal{\bm{w}}$ denote the initial peculiar acceleration
field constrained by 
\[
\nabla_X\!
\cdot{}^{(N)}\!\mal{\bm{w}}=-(1+z_I)\Delta_X \mal{\phi}
=-2/(3t_I^2(1+z_I))\lambda \mal{\delta}
\]
because of $\mal{a}=1/(1+z_I)$ where $z_I$ denotes the redshift 
of the initial time.
With a superposition 
ansatz for the Lagrangian perturbations of an Einstein-de 
Sitter background we have 
obtained the following family of trajectories ${}^{(N)}\!\bm{f}$
as an irrotational solution of  
Eq.(\ref{eqn:dn}) up to the second order: 

\begin{eqnarray}
&&q_z=\frac{3}{2}\left[\left(\frac{t}{t_I}\right)^{2/3}-1\right],  
\nonumber \\
&&q_{zz}=\left( \frac{3}{2}\right)^2 \left[-\frac{3}{14}
\left(\frac{t}{t_I}\right)^{4/3}
 +\frac{3}{5}\left(\frac{t}{t_I}\right)^{2/3}-\frac{1}{2}+\frac{4}
{35}\left(\frac{t}{t_I}\right)^{-1}\right], \label{eqn:ec}
\end{eqnarray}
and 
\begin{eqnarray}
&&\Delta_X \varphi^{(1)}=-\frac{2}{3}\mal{\delta}, \label{eqn:nden}\\
&&\Delta_X \varphi^{(2)}=\varphi^{(1)}_{|ii}\varphi^{(1)}_{|jj}
 -\varphi^{(1)}_{ij}\varphi^{(1)}_{ij}.
\end{eqnarray}
 Note that $\mal{\phi}$ is the following  order:
\begin{equation}
\mal{\phi}(\bm{X})=-\lambda \frac{\varphi^{(1)}}{t_I^2(1+z_I)^2}.
\label{eqn:za}
\end{equation}

Let us now turn to the longitudinal solution for the PN displacement vector. 
First of all, we need to express the Newtonian 
potential $\phi_N$ in terms of the known  trajectory field 
${}^{(N)}\!\bm{f}$ because Eq.(\ref{eqn:dj}) for solving the
PN displacement vector includes $\phi_N$ in the 
source terms.
 Using  Eq.(\ref{eqn:dm}),  Eq.(\ref{eqn:df}) can be rewritten as 
\[
{}^{(N)}\!h_{j,i} \phi_{N|j}=
\frac{d}{dt}\left[a^2\frac{d{}^{(N)}\!f^i}{dt}\right],\nonumber 
\]
and multiplying ${}^{(N)}\!f_{i|k}$ which is inverse of 
${}^{(N)}\!h_{j,i}$, we get 
\begin{equation}
\phi_{N|k}={}^{(N)}\!f_{i|k}\frac{d}{dt}\left[a^2
\frac{d{}^{(N)}\!f^i}{dt}\right].
\label{eqn:do}
\end{equation}
Moreover, by differentiating the both sides of Eq.(\ref{eqn:do}), 
we can get the Poisson's equation with respect to 
the Lagrangian coordinates:
\begin{eqnarray}
&&\Delta_X \phi_N=-\lambda \left( 2a \dot{a} 
\dot{q}_z +a^2\ddot{q}_z\right)
\Delta_X\varphi^{(1)}-\lambda^2\left[ \left( 2a \dot{a} \dot{q}_{zz}
 +a^2\ddot{q}_{zz}\right)
\Delta_X\varphi^{(2)}+q_z\left( 2a \dot{a} \dot{q}_{z} 
+a^2\ddot{q}_{z}\right)\left(
\varphi^{(1)}_{|ij}\varphi^{(1)}_{|i}\right)_{|j}\right] +O(\lambda^3)
\end{eqnarray}
Therefore we can arrive at the expression of $\phi_N$ with respect
to the Lagrangian coordinates:
\begin{eqnarray}
&&\phi_N(\bm{X},t)=-\lambda \left( 2a \dot{a} \dot{q}_z 
+a^2\ddot{q}_z\right)
\varphi^{(1)}(\bm{X})-\lambda^2 \left[\left( 2a \dot{a} \dot{q}_{zz} 
+a^2\ddot{q}_{zz}\right)\varphi^{(2)}(\bm{X})
+q_z\left( 2a \dot{a} \dot{q}_z +a^2\ddot{q}_z\right)
\tilde{\phi}(\bm{X})\right]+O(\lambda^3)
\nonumber \\
&&\hspace{4em}=-\lambda \frac{1}{t_I^2(1+z_I)^2}\varphi^{(1)}
-\lambda^2\left[ \left( 2a \dot{a} \dot{q}_{zz} 
+a^2\ddot{q}_{zz}\right)
\varphi^{(2)}+q_z\left( 2a \dot{a} \dot{q}_{z} 
+a^2\ddot{q}_{z}\right)\tilde{\phi}
(\bm{X})\right] +O(\lambda^3),\label{eqn:newtonpot}
\end{eqnarray}
where 
\begin{equation}
\tilde{\phi}(\bm{X})\equiv- \int \!d\bm{Y}\frac{
\left(\varphi^{(1)}_{|ij}(\bm{Y})
\varphi^{(1)}_{|i}(\bm{Y})\right)_{|j}}{4\pi |\bm{X}-\bm{Y}|}.
\end{equation}
Here it is easy to check that 
the Newtonian potential $\phi_N$ satisfies the 
initial condition Eq.(\ref{eqn:za}).  
It should be noted that the order $\lambda$ part of $\phi_N$ becomes 
constant because of considering only the growing mode of $q_z$, 
which is consistent with the result of linear theory 
on an Einstein-de
Sitter background as  discussed in \S \ref{cosmopn}.

Now to get the evolution equation for ${}^{(PN)}S_{|i}$, we
consider  Eq.(\ref{eqn:dj}). Similarly we make the following 
longitudinal perturbation ansatz:
\begin{equation}
{}^{(PN)}\!S_{|i}=\lambda Q_z(t) \Phi^{(1)}_{|i}(\bm{X}) 
+\lambda^2 Q_{zz}(t) \Phi^{(2)}
_{|i}(\bm{X})
+O(\lambda^3).  
\end{equation}
We must consider the initial conditions of $Q_z \Phi^{(1)}$, $Q_{zz}
\Phi^{(2)}$ and
so on in order to solve the second rank differential equation (\ref{eqn:dj}) 
according to the above ansatz. Part of the initial condition must 
be imposed by the definition of the displacement vectors:
\begin{eqnarray}
&&Q_z(t_I)\Phi^{(1)}_{|i}(\bm{X})=0,\hspace{2em} Q_{zz}(t_I)
\Phi^{(2)}_{|i}(\bm{X})=0,
\cdots \label{eqn:zv}
\end{eqnarray}
Although we need another initial condition for the  
peculiar 
velocity field, 
we leave $\dot{Q}_z(t_I) \Phi^{(1)}$ and $\dot{Q}_{zz}(t_I) \Phi^{(2)}$
and so on arbitrary for the present and think about them later.

The order $\lambda ^1$ part of Eq.(\ref{eqn:dj}) becomes
\begin{equation}
\lambda \left[\ddot{ Q}_z+2\frac{\dot{a}}{a}\dot{Q}_z
-\frac{2}{3t^2}Q_z\right] \Delta_X
\Phi^{(1)}(\bm{X})=\lambda \frac{2}{3t^2} \frac{ 
2 \varphi^{(1)}(\bm{X})}{(1+z_I)^2
t_I^2}.\label{eqn:dr}
\end{equation}
Hence we have
\begin{equation}
\ddot{Q}_z+2\frac{\dot{a}}{a}\dot{Q}_z-\frac{2}{3t^2}Q_z=\frac{1}{t^2} 
\label{eqn:zu}
\end{equation}
with
\begin{eqnarray}
\Delta_X \Phi^{(1)}(\bm{X})=-\frac{2}{3}2 \mal{\phi}(\bm{X})
=\frac{4}{3} \frac{\varphi^{(1)}(\bm{X})}{t_I^2(1+z_I)^2} .
\label{eqn:dq}
\end{eqnarray}
It may be noted here that $\Phi^{(1)}$ defined by Eq.(\ref{eqn:dq}) 
is the same as 
so-called ``superpotential'' on the hypersurface at the initial 
time $t_I$ \cite{ch}.     
We can find the general solution of  Eq.(\ref{eqn:zu}):  
\begin{equation}
Q_z=c_1t^{2/3}-\frac{3}{2}+c_2t^{-1} \label{eqn:eb},
\end{equation}
where $c_1$ and $c_2$ are   constants of integration. 
To determine them, we make use of the density filed (\ref{eqn:db})
solved up
to the first PN order, and consider the linear stage
where $q\varphi_{|i}$ and $Q\Phi_{|i}/c^2$ are much smaller than unity.
Then the linear order in $\lambda$ of the density field 
(\ref{eqn:db}) becomes 
\begin{eqnarray}
&&\delta(\bm{X},t)=\mal{\delta}(\bm{X})-q_z(t)\Delta_X\varphi^{(1)}(\bm{X})
-\frac{1}{c^2}Q_z(t)\Delta_X
\Phi^{(1)}(\bm{X})+O(\lambda^2),\hspace{2em}
\mbox{for}\hspace{1em}q_z(t)\Delta_X\!\varphi^{(1)},\frac{1}{c^2}
  Q_z(t)\Delta_X\Phi^{(1)}\ll 1\label{eqn:zw},
\end{eqnarray} 
where we used 
the fact that $\phi= \mal{\phi}$ in the order $\lambda^1$
through Eq.(\ref{eqn:newtonpot}). By
substituting Eqs.(\ref{eqn:ec}), (\ref{eqn:nden}),
(\ref{eqn:dq}) and (\ref{eqn:eb}) into Eq.(\ref{eqn:zw}), we obtain 
\begin{eqnarray}
\delta(\bm{X},t)=\left( \frac{t}{t_I}\right)^{2/3}
\mal{\delta}(\bm{X})+\frac{4}{3c^2}
\mal{\phi}(\bm{X})\left(c_1t^{2/3}-\frac{3}{2}+c_2 t^{-1}\right).
\end{eqnarray}
By comparing this with the density contrast field (\ref{eqn:zt}) derived in 
the linearized theory within the framework of our PN formalism, we can 
arrive at the solution of $Q_z$:
\begin{equation}
Q_z(t)=q_z(t)=\left(\frac{3}{2}\right)
\left[\left(\frac{t}{t_I}\right)^{2/3}-1\right].\label{eqn:zx}
\end{equation}
It should be again remarked that although the density contrast is supposed 
to be much smaller than unity in linearized theory, 
we do not have to assume such smallness for the
density contrast in Lagrangian perturbation theory.    

The initial conditions for the quantities of higher order in $\lambda$ 
has no corresponding counterparts in linearized theory and thus 
we are not able to know the initial conditions for such quantities. 
We may, however, expect them to be much smaller than the order 
$\lambda$ quantities at initial time so that we will neglect them and 
we adopt the following 
initial condition for $Q_{zz}\Phi^{(2)}_{|i}$ and so on:
\begin{equation}
\dot{Q}_{zz}(t_I)\Phi^{(2)}_{|i}(\bm{X})=0,\cdots.   \label{eqn:ds}
\end{equation}
We are now in the position to be able 
to solve $ Q_{zz}\Phi^{(2)}(\bm{X})$. The differential
equation for $Q_{zz}\Phi^{(2)}$ becomes the following form 
by using the fact that $2a\dot{a}\dot{q}_z+a^2
\ddot{q}_z=1/((1+z_I)^2t_I^2)$ 
\begin{eqnarray}
\lefteqn{\left(\ddot{Q}_{zz}+2 \frac{\dot{a}}{a}\dot{Q}_{zz}
-\frac{2}{3t^2}Q_{zz}\right) 
\Delta_X \Phi^{(2)}(\bm{X})} \nonumber \\
&&\hspace{1em}=\left[-q_z\left(\ddot{Q}_z
+2\frac{\dot{a}}{a}\dot{Q}_z\right)
-Q_z\left(\ddot{q}_z+2\frac{\dot{a}}{a}\dot{q}_z\right)
+\frac{2}{3t^2}Q_zq_z\right]
\left(\Phi^{(1)}_{|ii}\varphi^{(1)}_{|jj}-\Phi^{(1)}_{|ij}
 \varphi^{(1)}_{|ij}\right)(\bm{X})
\nonumber \\
&&\hspace{2em}+\left[-3\left\{\left(6\frac{\dot{a}^3}{a}
+12\dot{a}\ddot{a}+2a
\frac{d^3a_{}}{dt^3}
\right)\dot{q}_{zz}+6\left(3\dot{a}^2+a\ddot{a}\right)\ddot{q}_{zz}
+9a\dot{a}\frac{d^3q_{zz}}{dt^3}+a^2
\frac{d^4q_{zz}}{dt^4}\right\}
+\frac{10}{3t^2}\left(2a\dot{a}\dot{q}_{zz}+a^2
\ddot{q}_{zz}\right)\right]\varphi^{(2)}(\bm{X}) \nonumber \\
&&\hspace{2em}+\left[-3\left(\ddot{q}_z+3
\frac{\dot{a}}{a}\dot{q}_z\right)\left(
2a\dot{a}\dot{q}_z+a^2\ddot{q}_z\right)
+\frac{10}{3t^2}q_z\left(2a\dot{a}\dot{q}_z
+a^2 \ddot{q}_z\right)\right]\tilde{\phi}(\bm{X})\nonumber \\
&&\hspace{2em}+\left[-\frac{1}{a^2}
\left(2a\dot{a}\dot{q}_z+a^2\ddot{q}_z\right)^2\!
-\frac{\dot{a}}{a}\dot{q}_z\left(2a\dot{a}\dot{q}_z
+a^2\ddot{q}_z\right)+\frac{2}{3t^2}
\left(-\frac{1}{2(1+z_I)^2t_I^2}
+\frac{5}{2}a^2\dot{q}_z^2\right)\right]\!
\left(\varphi^{(1)}_{|i}(\bm{X})\right)^2\nonumber \\
&&\hspace{2em}+\frac{2}{3t^2}2\mal{\delta}(\bm{X})\frac{\varphi^{(1)}(\bm{X})}
{(1+z_I)^2t_I^2}.
\label{eqn:ea}
\end{eqnarray}  
By using Eqs.(\ref{eqn:ec}) and (\ref{eqn:zx}), this 
equation is rewritten in the following simple form: 
\begin{eqnarray}
\lefteqn{\left(\ddot{Q}_{zz}+2\frac{\dot{a}}{a}\dot{Q}_z
-\frac{2}{3t^2}Q_{zz}\right)
\Delta_X\Phi^{(2)}(\bm{X})} \nonumber \\
&&=\frac{1}{t_I^2}\left[-\frac{3}{5}\left(\frac{t}{t_I}\right)^{-2/3}+
\left(\frac{t}{t_I}\right)^{-2}-\frac{2}{5}\left(
\frac{t}{t_I}\right)^{-7/3}
\right]\left(\Phi^{(1)}_{|ii}\varphi^{(1)}_{|jj}
-\Phi^{(1)}_{|ij}\varphi^{(1)}_{|ij}\right)
(\bm{X}) \nonumber \\
&&\hspace{1em}+\frac{2}{3t_I^2}\left(\frac{t}{t_I}\right)^{-2}
\frac{1}{(1+z_I)^2t_I^2}\left[2\mal{\delta}(\bm{X})\varphi^{(1)}(\bm{X})
-\frac{1}{2}\left(\varphi^{(1)}_{|i}(\bm{X})\right)^2
-\frac{15}{2}\tilde{\phi}(\bm{X})
+\frac{9}{2}\varphi^{(2)}(\bm{X})\right].\label{eqn:ed}
\end{eqnarray}
To obtain solutions of Eq.(\ref{eqn:ed}), we make use of the
linearity of Poisson's equation
 and split the potential $\Phi^{(2)}$into two parts.
Then, we  have to solve separately the 
linear ordinary differential equation:
\begin{equation}
\ddot{Q}^{(a)}_{zz}+2\frac{\dot{a}}{a}\dot{Q}^{(a)}_{zz}
-\frac{2}{3t^2}Q^{(a)}_{zz}
=\frac{1}{2t_I^2}\left[-\frac{3}{5}\left(\frac{t}{t_I}\right)^{-2/3}
+\left(\frac{t}{t_I}\right)^{-2}-\frac{2}{5}\left(
\frac{t}{t_I}\right)^{-7/3}\right],
\end{equation}
with
\begin{equation}
\Delta_X\Phi^{(2a)}=2\left(\Phi^{(1)}_{|ii}\varphi^{(1)}_{|jj}
-\Phi^{(1)}_{|ij}\varphi^{(1)}_{|ij}\right),
\end{equation}
and 
\begin{equation}
\ddot{Q}^{(b)}_{zz}+2\frac{\dot{a}}{a}\dot{Q}^{(b)}_{zz}
-\frac{2}{3t^2}Q^{(b)}_{zz}
=\frac{2}{3t_I^2}\left(\frac{t}{t_I}\right)^{-2},
\end{equation}
with
\begin{equation}
\Delta_X\Phi^{(2b)}=\frac{1}{(1+z_I)^2t_I^2}\left[
2\mal{\delta}\varphi^{(1)}-\frac{1}{2}
\left(\varphi^{(1)}_{|i}\right)^2-\frac{15}{2}\tilde{\phi}
+\frac{9}{2}\varphi^{(2)}\right].
\end{equation}
We find the following solution of $Q_{zz}$
for the initial conditions Eqs.(\ref{eqn:zv}) and (\ref{eqn:ds}):
\begin{eqnarray}
&&Q_{zz}^{(a)}=q_{zz}(t)=\left(\frac{3}{2}\right)^2
\left[-\frac{3}{14}\left(\frac{t}{t_I}\right)^{4/3}+\frac{3}{5}
\left(\frac{t}{t_I}\right)^{2/3}-\frac{1}{2}+\frac{4}{35}\left(\frac{t}{t_I}
\right)^{-1}\right],  \\
&&Q_{zz}^{(b)}=\left(\frac{3}{2}\right)
\left[\frac{2}{5}\left(\frac{t}{t_I}\right)^{2/3}
-\frac{2}{3}+\frac{4}{15}\left(\frac{t}{t_I}\right)^{-1}\right].
\end{eqnarray}
We have obtained the following family of trajectories $\bm{x}
=\bm{f}(\bm{X},t)$ as an 
irrotational solution up to the second order of $\lambda$ in  
the perturbed  Einstein-de Sitter background:
{
\setcounter{enumi}{\value{equation}}
\addtocounter{enumi}{1}
\setcounter{equation}{0}
\renewcommand{\theequation}{\thesection.\theenumi\alph{equation}}
\begin{eqnarray}
&&\bm{f}(\bm{X},t)
=\bm{X}
+q_z(t)  \nabla_X \left( \varphi^{(1)}(\bm{X})
     +{1\over c^2}\Phi^{(1)}(\bm{X}) \right)
+q_{zz}(t)  \nabla_X \!\left( \varphi^{(2)}(\bm{X}) 
     +{1\over c^2} \Phi^{(2a)}(\bm{X}) \right)
 +\frac{1}{c^2}Q_{zz}^{(b)}(t)\nabla_X\Phi^{(2b)}(\bm{X}),
\label{eqn:ee}        
\end{eqnarray}
with
\begin{eqnarray}
&&q_z=\left(\frac{3}{2}\right)\left[\left(
\frac{t}{t_I}\right)^{2/3}-1\right], \\
&&q_{zz}=\left(\frac{3}{2}\right)^2
\left[-\frac{3}{14}\left(\frac{t}{t_I}\right)^{4/3}+\frac{3}{5}
\left(\frac{t}{t_I}\right)^{2/3}-\frac{1}{2}+\frac{4}{35}\left(\frac{t}{t_I}
\right)^{-1}\right], \\
&&Q_{zz}^{(b)}=\left(\frac{3}{2}\right)
\left[\frac{2}{5}\left(\frac{t}{t_I}\right)^{2/3}
-\frac{2}{3}+\frac{4}{15}\left(\frac{t}{t_I}\right)^{-1}\right],
\end{eqnarray}
and
\begin{eqnarray}
&&\Delta_X \varphi^{(1)}=-\frac{2}{3}\mal{\delta}, \\
&&\Delta_X \varphi^{(2)}=\varphi^{(1)}_{|ii}\varphi^{(1)}_{|jj}
-\varphi^{(1)}_{|ij}\varphi^{(1)}_{|ij}, \\
&&\Delta_X \Phi^{(1)}=-\frac{4}{3}\mal{\phi}
=\frac{4}{3(1+z_I)^2t_I^2}\varphi^{(1)}, \\
&&\Delta_X \Phi^{(2a)}=2\left(\Phi^{(1)}_{|ii}\varphi^{(1)}_{|jj}-
\Phi^{(1)}_{|ij}\varphi^{(1)}_{|ij}\right), \\
&&\Delta_X \Phi^{(2b)}=\frac{1}{(1+z_I)^2t_I^2}\left[
2\mal{\delta}\varphi^{(1)}-\frac{1}{2}
\left(\varphi^{(1)}_{|i}\right)^2-\frac{15}{2}\tilde{\phi}
+\frac{9}{2}\varphi^{(2)}\right].
\end{eqnarray}\setcounter{equation}{\value{enumi}}}\noindent

\subsubsection{Physical Interpretation of the longitudinal perturbation
  solution}

In this subsection,
 we will express  the trajectory field (\ref{eqn:ee}) 
in terms of  physical quantities.
Our coordinate conditions (\ref{eqn:a}) show that the density contrast
field $\delta$ 
has gauge ambiguities. 
In order to interpret the trajectory field (\ref{eqn:ee}) physically, 
we should express $f^i(\bm{X},t)$ in terms of the gauge invariant 
quantities at the initial time, rather than the coordinate dependent
initial density field 
$ \mal{\delta} $, because  only gauge invariant quantities are physically 
relevant. This will turn out to be essential for the physical interpretation. 
The physical density contrast becomes at the initial time $t_I$ 
\begin{equation}
 \mal{\delta}_g(\bm{X})\equiv\mal{\delta}+\frac{2}{c^2}\mal{\alpha}^{(2)}\simeq
 \mal{\delta}(\bm{X})+\frac{2}{c^2}\mal{\phi}(\bm{X}), \label{eqn:zz}
\end{equation}  
where
\[
\Delta_X\mal{\alpha}^{(2)}\!\!(\bm{X})\equiv 4\pi G \mal{a}^2\mal{\rho_b}
\mal{\delta_g}(\bm{X}) \nonumber 
\]
and $\mal{\delta}_g$ denotes the gauge invariant quantity at the initial 
time defined by
linearized theory, and $\alpha^{(2)}/c^2=\alpha-1$. It is supposed
that $\alpha^{(2)}$ remains constant
with respect to  time in the linear regime because of neglecting the  
decaying mode. (The reason for this choice is described in Appendix A.) 
Thus, the quantity 
$\mal{\delta}$ has the spurious mode $2\mal{\phi}/c^2$ caused by
our gauge conditions (\ref{eqn:a}), so we had better redefine the  
displacement vector of (\ref{eqn:ee}) in terms of $\delta_g$, that is, since
we have
\[
 \Delta_X\left(\varphi^{(1)}+\frac{1}{c^2}\Phi^{(1)}(\bm{X})\right)=
-\frac{2}{3}\left(\mal{\delta}(\bm{X})+\frac{2}{c^2}\mal{\phi}(\bm{X})\right),
\]
we may redefine the first  order velocity potential in Newtonian theory as 
\begin{equation}
\Delta_X\Psi_N^{(1)}\equiv -\frac{2}{3}\mal{\delta}_g(\bm{X}).
\end{equation}
Naturally, the quantity $\Psi_N$ is a gauge invariant variable. 
Thus the lowest  PN irrotational velocity potential 
 $\Phi^{(1)}$ in our coordinates 
is a spurious mode generated by the gauge ambiguities of 
the density contrast $\mal{\delta}$ in our coordinates. 
Therefore, 
it has no physical meaning  and  
 can be understood as a part of the gauge invariant 
Newtonian first order quantities.  
Likewise since we have
\begin{eqnarray}
\Delta_X\left(\varphi^{(2)}(\bm{X})+\frac{1}{c^2}\Phi^{(2a)}\right)&=&
\varphi^{(1)}_{ii}\varphi^{(1)}_{jj}-\varphi^{(1)}_{ij}\varphi^{(1)}_{ij}
+{2\over c^2}\left(\varphi^{(1)}_{ii}
\Phi^{(1)}_{jj}-\varphi^{(1)}_{ij}\Phi^{(1)}_{ij}\right) \nonumber \\
&=&\left(\Psi^{(1)}_{N|ii}\Psi^{(1)}_{N|jj}-\Psi^{(1)}_{N|ij}\Psi^{(1)}_{N|ij}\right)
+O(c^{-4}), \nonumber  
\end{eqnarray}
we may redefine the second  order potential in Newtonian theory as
\begin{equation} 
\Delta_X\Psi^{(2)}_N\equiv \Psi^{(1)}_{N|ii}\Psi^{(1)}_{N|jj}
-\Psi^{(1)}_{N|ij}\Psi^{(1)}_{N|ij}.
\end{equation}
Furthermore for the first PN displacement vector $\Phi^{(2b)}$ we have
\begin{eqnarray}
\Delta_X\Phi^{(2b)}&=&\!\!\frac{1}{(1+z_I)^2t_I^2}\left[
2\mal{\delta}\varphi^{(1)}-\frac{1}{2}\left(\varphi^{(1)}_{|i}\right)^2+\frac{15}{2}
\int\!d\bm{Y}\frac{\left(\varphi^{(1)}_{|ij}(\bm{Y})\varphi^{(1)}_{|i}(\bm{Y})
\right)_{|j}}{4\pi |\bm{X}-\bm{Y}|}+\frac{9}{2}\varphi^{(2)}\right] \nonumber \\
&=&\!\!\frac{1}{(1+z_I)^2t_I^2}\left[
2\mal{\delta}_g\Psi^{(1)}_N\!-\!\frac{1}{2}\left(\Psi^{(1)}_{N|i}\right)^2
+\!\frac{15}{2}
\int\!d\bm{Y}\frac{\left(\Psi^{(1)}_{N|ij}(\bm{Y})\Psi^{(1)}_{N|i}(\bm{Y})
\right)_{|j}}{4\pi |\bm{X}-\bm{Y}|}\!+\!\frac{9}{2}\Psi_N^{(2)}\right]\!+\!O(c^{-4}), \nonumber
\end{eqnarray}
so we may redefine the first  order potential in the fist PN approximation as
\begin{eqnarray}
&&\Delta_X\Psi_{PN}\equiv \frac{1}{(1+z_I)^2t_I^2}\left[2\mal{\delta}_g\Psi^{(1)}_N\!
-\!\frac{1}{2}\left(\Psi^{(1)}_{N|i}\right)^2+\!\frac{15}{2}
\int\!d\bm{Y}\frac{\left(\Psi^{(1)}_{N|ij}(\bm{Y})\Psi^{(1)}_{N|i}(\bm{Y})
\right)_{|j}}{4\pi |\bm{X}-\bm{Y}|}\!+\!\frac{9}{2}\Psi_N^{(2)}\right].
\end{eqnarray}

In the end, we have obtained the following family of trajectories $\bm{x}=\bm{f}(\bm{X},t)$
as a physical 1st PN irrotational  solution up to the second order of $\lambda$  in the
perturbed Einstein-de Sitter background:
{
\setcounter{enumi}{\value{equation}}
\addtocounter{enumi}{1}
\setcounter{equation}{0}
\renewcommand{\theequation}{\thesection.\theenumi\alph{equation}}
\begin{eqnarray}
&&\bm{f}(\bm{X},t)
=\bm{X}
+q_z(t)  \nabla_X \Psi^{(1)}_N(\bm{X}) 
+q_{zz}(t)  \nabla_X \Psi^{(2)}_N(\bm{X})
+\frac{1}{c^2}Q_{zz}^{(b)}(t)\nabla_X\Psi_{PN}(\bm{X}), 
\label{eqn:ww}
\end{eqnarray}
with 
\begin{eqnarray}
&&q_z=\left(\frac{3}{2}\right)\left[\left(\frac{t}{t_I}\right)^{2/3}-1\right],\\
&&q_{zz}=\left(\frac{3}{2}\right)^2\left[-\frac{3}{14}\left(\frac{t}{t_I}\right)^{4/3}
+\frac{3}{5}\left(\frac{t}{t_I}\right)^{2/3}-\frac{1}{2}+\frac{4}{35}\left(
\frac{t}{t_I}\right)^{-1}\right],\\
&&Q_{zz}^{(b)}=\left(\frac{3}{2}\right)\left[\frac{2}{5}\left(\frac{t}{t_I}\right)^{2/3}
-\frac{2}{3}+\frac{4}{15}\left(\frac{t}{t_I}\right)^{-1}\right],
\end{eqnarray}
and
\begin{eqnarray}
&&\Delta_X\Psi^{(1)}_N=-\frac{2}{3}\mal{\delta}_g,\\
&&\Delta_X\Psi^{(2)}_N=\Psi^{(1)}_{N|ii}\Psi^{(1)}_{N|jj}
-\Psi^{(1)}_{N|ij}\Psi^{(1)}_{N|ij},\label{eqn:newvelsecond}\\
&&\Delta_X\Psi_{PN}=\frac{1}{(1+z_I)^2t_I^2}\!\!
\left[2\mal{\delta}_g\Psi^{(1)}_N\!-\!\frac{1}{2}\left(\Psi^{(1)}_{N|i}\right)^2
+\!\frac{15}{2}
\int\!d\bm{Y}\frac{\left(\Psi^{(1)}_{N|ij}(\bm{Y})\Psi^{(1)}_{N|i}(\bm{Y})
\right)_{|j}}{4\pi |\bm{X}-\bm{Y}|}\!+\!\frac{9}{2}\Psi_N^{(2)}\right]\!.
\end{eqnarray}\setcounter{equation}{\value{enumi}}}\noindent

It should be again noted that  Eq.(\ref{eqn:ww}) is expressed only 
in terms of  the gauge invariant quantities at initial time when  the linearized 
approximation is valid. 
In other words, if using  trajectory field (\ref{eqn:ww}), we 
can accurately describe evolutions of 
all scale fluctuations in the {\it early} stage 
according to the gauge invariant
linearized theory, and also take into account   
the relativistic 
PN corrections which are generated by the non-linearities due to
evolutions of fluctuations 
in the {\it subsequent}  structure formation (see below in detail).     

\subsection{The transverse perturbation solution at 1st PN order}

We now consider the transverse part of the trajectory field at  PN
order. We assumed that the vorticity for the Newtonian displacement
vector is negligible, which seems to be 
reasonable \cite{buchert92,buchert94,peebles}.
In this section, it will be found that the PN displacement vector 
has a transverse flow with the growing mode in the Lagrangian
coordinates even in the case.

According to the result (\ref{eqn:ww}) in the previous section, 
we consider the following ansatz for the transverse perturbations 
superposed on the solved irrotational 
trajectory field up to the first PN order:
\begin{eqnarray}
f^i(\bm{X},t)&=&{}^{(N)}\!f^i(\bm{X},t)+\lambda^2\frac{1}{c^2}
Q^{(b)}_{zz}(t)\Psi_{PN|i}(\bm{X})+\frac{1}{c^2}
\left(\lambda {}^{(PN)}\!S^{(1)T}_i(\bm{X},t)+\lambda^2
{}^{(PN)}\!S^{(2)T}_i\left(\bm{X},t\right)
\right),
\label{eqn:transtraj}
\end{eqnarray}
where 
\[
{}^{(N)}\!f^i(\bm{X},t)=X^i+\lambda q_z(t)\Psi^{(1)}_{N|i}(\bm{X})
+\lambda^2q_{zz}(t)\Psi^{(2)}_{N|i},
\]
and the PN transverse displacement vector at each order is constrained by,
\begin{equation}
\nabla_X\cdot {}^{(PN)}\!\bm{S}^{(1)T}=\nabla_X\cdot
 {}^{(PN)}\!\bm{S}^{(2)T}=
 0.
\end{equation}
We need only Eq.(\ref{eqn:transtrajdiff}) derived in
\S \ref{rellagevo} to solve the transverse part 
${}^{(PN)}S^{Ti}$ 
of the
PN displacement vector:
{
\setcounter{enumi}{\value{equation}}
\addtocounter{enumi}{1}
\setcounter{equation}{0}
\renewcommand{\theequation}{\ref{eqn:transtrajdiff}}
\begin{eqnarray}
\lefteqn{\epsilon_{abc}f_{j|a}f_{i|b}\left[\frac{d^2 f_{j|c}}{dt^2}+
2\frac{\dot{a}}{a}\frac{df_{j|c}}{dt}\right]}\nonumber \\
&&=
-\frac{J}{c^2}\left[\frac{1}{a^2}\frac{\partial }{\partial t}
\!\left(a^2\epsilon_{ijk}\beta^{(3)k}{}_{,j}\right)
+\epsilon_{ijk}\frac{\partial }{\partial x^j}\!
\left\{v^l\left(\beta^{(3)k}{}_{,l}-\beta^{(3)l}{}_{,k}\right)\right\}
\right.\nonumber \\
&&\hspace{1em}\left.+\frac{2}{J}\epsilon_{abc}f_{i|b}\frac{df_l}{dt}
\frac{df_{l|c}}{dt}\phi_{|a}\right]-\frac{1}{c^2}\epsilon_{abc}f_{j|a}
f_{i|b}\left(\frac{df_{j|c}}{dt}\frac{d A}{dt}+\frac{df_j}{dt}
\frac{dA_{|c}}{dt}\right)+O\!\left(c^{-4}\right),  
\end{eqnarray}\setcounter{equation}{\value{enumi}}}\noindent
Naturally, the order $c^0$ in this equation
produces Newtonian  counterparts 
in the case that the trajectory field is introduced on the comoving
coordinates \cite{buchert92,buchert93,buchert94}. 
Then taking into account  order of $\lambda$ in each 
term of the right-hand side and using the assumption (\ref{eqn:deterassume}),
the above equation becomes  
\begin{eqnarray}
&&\epsilon_{abc}f_{j|a}f_{i|b}\left[ \ddot{f}_{j|c}+
2\frac{\dot{a}}{a}\dot{f}_{j|c}\right]
=-\frac{J_N}{c^2}\frac{1}{a^2}\frac{\partial }{\partial t}
\!\left(a^2\epsilon_{ijk}\beta^{(3)k}{}_{,j}\right)
-\frac{J_N}{c^2}\epsilon_{ijk}\frac{\partial}{\partial x^j}\!
  \left[{}^{(N)}\!\dot{f}^l(\beta^{(3)k},_l-\beta^{(3)l},_k)\right]\!\!
  +O\!\left(c^{-4},\lambda^3\right), \label{eqn:transeqns}
\end{eqnarray}
where 
we neglected the terms like $\phi_N v^2$ in our approximation 
because they are higher than third order $\lambda^3$ as derived in 
the previous section. 
This equation shows that  the shift vector generates the first PN transverse part. 
Inserting the ansatz (\ref{eqn:transtraj}),   
we can rewrite the left-hand side of the above equation
(\ref{eqn:transeqns})
  as  
\begin{eqnarray}
 \epsilon_{abc}f_{j|a}f_{i|b}\left[ \ddot{f}_{j|c}+
2\frac{\dot{a}}{a}\dot{f}_{j|c}\right] \nonumber \\
&&\hspace{-10em}=\frac{1}{c^2}\left[\lambda\epsilon_{ijk}
\left({}^{(PN)}\!
\ddot{S}^{(1)T}_{k|j} 
+2\frac{\dot{a}}{a}{}^{(PN)}\!\dot{S}^{(1)T}_{k|j}\right)
+\lambda^2\epsilon_{ijk}\left({}^{(PN)}\!\ddot{S}^{(2)T}_{k|j}
+2\frac{\dot{a}}{a} 
 {}^{(PN)}\!S^{(2)T}_{k|j}\right)
+\lambda^2\left(\ddot{q}_z+2\frac{\dot{a}}{a}
\dot{q}_{z}\right)\epsilon_{ijk}\Psi^{(1)}_{N|jl}{}^{(PN)}\!S^{(1)T}_{l|k}
\right.  \nonumber \\
&&\hspace{-9em}\left.+\lambda^2 q_z\left(\epsilon_{ijk}\Psi^{(1)}_{N|kl}
-\epsilon_{jkl}\Psi^{(1)}_{N|ik}\right)\left(
{}^{(PN)}\!\ddot{S}^{(1)T}_{l|j}+2\frac{\dot{a}}{a}{}^{(PN)}\!\dot{S}^{(1)T} 
_{l|j}
\right)\right]
+O\!\left(c^{-4},
\lambda^3\right).\nonumber 
\end{eqnarray}
Accordingly, as the irrotational case, we can obtain the following
differentiation equation for solving the
transverse part of the first PN displacement vector
at the order $c^{-2}$ of the both side in the above equation:
\begin{eqnarray}
&&\lambda\epsilon_{ijk}
\left({}^{(PN)}\!
\ddot{S}^{(1)T}_{k|j} 
+2\frac{\dot{a}}{a}{}^{(PN)}\!\dot{S}^{(1)T}_{k|j}\right)
+\lambda^2\epsilon_{ijk}\left({}^{(PN)}\!\ddot{S}^{(2)T}_{k|j}+2\frac{\dot{a}}{a} 
 {}^{(PN)}\!S^{(2)T}_{k|j}\right)+\lambda^2\left(\ddot{q}_z+2\frac{\dot{a}}{a}
\dot{q}_{z}\right)\epsilon_{ijk}\Psi^{(1)}_{N|jl}{}^{(PN)}\!S^{(1)T}_{l|k}
  \nonumber \\
&&\hspace{14em}+\lambda^2 q_z\left(\epsilon_{ijk}\Psi^{(1)}_{N|kl}
-\epsilon_{jkl}\Psi^{(1)}_{N|ik}\right)\left(
{}^{(PN)}\!\ddot{S}^{(1)T}_{l|j}+2\frac{\dot{a}}{a}{}^{(PN)}\!\dot{S}^{(1)T} 
_{l|j}
\right) \nonumber \\
&&\hspace{4em}=
-\frac{J_N}{a^2}\frac{\partial }{\partial t}\left(a^2
  \epsilon_{ijk}\beta^{(3)k}{}_{,j}\right)
 -J_N\epsilon_{ijk}\frac{\partial}{\partial x^j}\!
  \left[v_N^l(\beta^{(3)k},_l-\beta^{(3)l},_k)\right]\!\!
  +O\!\left(\lambda^3\right). \label{eqn:pntransdiff}
\end{eqnarray}
The form of this equation shows that it  could be solved 
if we  expressed the shift vector $\beta^{(3)i}$ in terms of
already-known Newtonian 
trajectory field ${}^{(N)}\!f^i$.

Therefore, we first have  to solve the   shift vector with respect to
the Lagrangian coordinates. 
Following the relativistic linearized theory \cite{bardeen,ks},
it will be convenient to decompose the shift vector $\beta^{(3)i}$
into its longitudinal part and its transverse part on 
the Einstein-de Sitter background:
\begin{equation}
 \beta^{(3)i}=\beta_{L,i}^{(3)}+\beta_T^{(3)i},\hspace{2em}
  \beta^{(3)i}_{T}{}_{,i}=0.\label{eqn:decomshift}
\end{equation}
 Eq.(\ref{eqn:pntransdiff}) shows that we need only the explicit form of
its transverse part $\beta^{(3)i}_T$ in the situation 
we are here interested in. 
Since the gauge condition (\ref{eqn:a}) implies that the transverse part
of shift vector is a gauge invariant quantity, 
any ambiguities caused by the gauge freedom do not remain.   
The perturbation quantity $\beta_T^{(3)i}$ is constrained by
Einstein equation (\ref{eqn:bm}):
\begin{eqnarray}
 \Delta_f\beta_T^{(3)i}
  =4\left(\frac{ \partial \phi_{N,i}}{\partial t}+
\frac {\dot{a}}{a}\phi_{N,i}\right)+16 \pi G a^2 \rho
   v_N^i.\label{eqn:shifttrans}
\end{eqnarray}
Since  the right-hand side of this equation is shown to be 
divergenceless exactly through the continuity equation
(\ref{eqn:bt}) at  Newtonian order, the condition (\ref{eqn:decomshift}) 
for $\beta^{(3)i}_T$ is always satisfied.  
We have already obtained the
expression of the peculiar velocity field and the peculiar gravitational
potential (\ref{eqn:map}) and (\ref{eqn:newtonpot}), respectively.
Therefore, it will be adequate to make use of the equation
(\ref{eqn:shifttrans}) for our purpose. 
 By using Eqs.(\ref{eqn:dm}) and (\ref{eqn:deterassume}),
we rewrite the left-hand side
of Eq.(\ref{eqn:shifttrans}) in
terms of the independent variables $\bm{X}$ and $t$:
\begin{eqnarray}
 \Delta_f\beta^{(3)i}_T&=&\frac{\partial X^k}{\partial
  x^j}\frac{\partial }{\partial X^k}\!\left[\frac{\partial X^l}{\partial 
  x^j}\frac{\partial }{\partial X^l}\!\beta^{(3)i}_T\right] \nonumber \\
 &=&{}^{(N)}\!h_{k,j}\frac{\partial }{\partial
  X^k}\!\left[{}^{(N)}\!h_{l,j}\beta^{(3)i}_{T}{}_{|l} \right]
   \nonumber \\
 &&\hspace{-2em}=\frac{1}{2J_N}\epsilon_{kab}\epsilon_{jcd}{}^{(N)}\!f_{c|a}
  {}^{(N)}\!f_{d|b}\frac{\partial }{\partial X^k}\!\left[
  \frac{1}{2J_N}\epsilon_{lpq}\epsilon_{jrs}{}^{(N)}\!f_{r|p}{}^{(N)}\!f_{s|q}
  \beta^{(3)i}_T{}_{|l}\right]\nonumber \\
 &&\hspace{-2em}=\frac{1}{2J_N^3}\epsilon_{kab}\epsilon_{lpq}{}^{(N)}\!f_{c|a}
  {}^{(N)}\!f_{d|b}\left[-J_{N|k}{}^{(N)}\!f_{c|p}{}^{(N)}\!f_{d|q}
  \beta^{(3)i}_T{}_{|l}+2J_N
  {}^{(N)}\!f_{c|pk}{}^{(N)}\!f_{d|q}\beta^{(3)i}_T{}_{|l}
+J_N{}^{(N)}\!f_{c|p}
  {}^{(N)}\!f_{d|q}\beta^{(3)i}_T{}_{|lk}\right] \nonumber \\
 &&\hspace{-2.5em}=\frac{1}{J_N^3}\left[\Delta_X\beta^{(3)i}_T-\lambda
  q_z\Psi^{(1)}_{N|jjk}\beta^{(3)i}_T{}_{|k}+3\lambda
  q_z\Psi^{(1)}_{N|jj}\beta^{(3)i}_T{}_{|kk}-2\lambda
  q_z\Psi^{(1)}_{N|jk}\beta^{(3)i}_T{}_{|jk}+O\!
  \left(\lambda^3\right)\right],\nonumber \\ 
\end{eqnarray}
where we used the fact that the perturbation quantity $\beta^{(3)i}_T$ 
is of order $\lambda^1$ at most and 
\[
J_N=1+\lambda q_z\Psi^{(1)}_{N|ii}+O\!\left(\lambda^2\right).
\]
Using Eqs.(\ref{eqn:db}), (\ref{eqn:df}) and (\ref{eqn:ww}),
 the right-hand 
side of Eq.(\ref{eqn:shifttrans}) can also be rewritten as 
\begin{eqnarray}
\mbox{(r.h.s)}&=&4J_N^3\left[\left(\frac{d}{dt}
-v_N^j\frac{\partial}{\partial x^j}\right)
\phi_{N,i}+\frac{\dot{a}}{a}\phi_{N,i} \right]
+\frac{16\pi G}{a} \mal{\rho}\mal{a}^3J_N^2v_N^i \nonumber \\
&&\hspace{0em}=-4J_N^3\left[\left(\frac{d}{dt}-{}^{(N)}\!\dot{f}_j 
\frac{\partial }{\partial x^j}\right)\left(a^2{}^{(N)}\!\ddot{f}_i+
2a\dot{a}{}^{(N)}\!\dot{f}_i\right)+\frac{\dot{a}}{a}\left(a^2{}^{(N)}
\!\ddot{f}_i+2a\dot{a}{}^{(N)}\!\dot{f}_i\right)\right]
+\frac{16\pi G }{a}\mal{\rho}\mal{a}^3{}^{(N)}\!\dot{f}_i 
\nonumber \\
&&\hspace{0em}=-4J_N^3\left[a^2\frac{d^3 \!{}^{(N)}\!f_i}{dt^3}+
5a\dot{a}{}^{(N)}\!\ddot{f}_i+2a^2\left(\frac{\dot{a}^2}{a^2}
+\frac{\ddot{a}}{a}\right){}^{(N)}\!\dot{f}_i
-\frac{1}{2J_N}\epsilon_{kab}
\epsilon_{jcd}{}^{(N)}\!f_{c|a}{}^{(N)}\!f_{d|b}\left(a^2{}^{(N)}\!
\ddot{f}_{i|k}
+2a\dot{a}{}^{(N)}\!\dot{f}_{i|k}\right){}^{(N)}\!\dot{f}_j\right]\nonumber \\
&&\hspace{2em}
+\frac{16\pi G }{a}\mal{\rho}_b\mal{a}^3(1+\lambda \mal{\delta}){}^{(N)}\!
\dot{f}_i \nonumber \\
&&\hspace{0em}=\lambda^2 
\frac{4}{t_I^3(1+z_I)^2}\left(\frac{t}{t_I}\right)^{-1/3}
\left(-\Psi^{(1)}_{N|jj}\Psi^{(1)}_{N|i}+\Psi^{(1)}_{N|ij}\Psi^{(1)}_{N|j}
+\Psi^{(2)}_{N|i}\right)+O\!\left(\lambda^3\right).
\end{eqnarray}
Finally, the equation (\ref{eqn:shifttrans}) becomes the following 
simple form in terms of the Lagrangian coordinates:
\begin{eqnarray}
&&\Delta_X\beta^{(3)i}_T-\lambda
  q_z\Psi^{(1)}_{N|jjk}\beta^{(3)i}_T{}_{|k}+3\lambda
  q_z\Psi^{(1)}_{N|jj}\beta^{(3)i}_T{}_{|kk}-2\lambda
  q_z\Psi^{(1)}_{N|jk}\beta^{(3)i}_T{}_{|jk}\nonumber \\
&&\hspace{1.5em}=\lambda^2 
\frac{4}{t_I^3(1+z_I)^2}\left(\frac{t}{t_I}\right)^{-1/3}
\left(-\Psi^{(1)}_{N|jj}\Psi^{(1)}_{N|i}+\Psi^{(1)}_{N|ij}\Psi^{(1)}_{N|j}
+\Psi^{(2)}_{N|i}\right)+O\!\left(\lambda^3\right).\label{eqn:shiftdiffer}
\end{eqnarray}
We are now in the position to be able to solve the shift vector  iteratively. 
The lowest order part of this equation yields 
\begin{equation}
\Delta_X\beta^{(3)i}_T=0, 
\end{equation}
so the solution of the shift vector  can be chosen as  
\[
\beta^{(3)i}_T=0+O\!\left(\lambda^2\right). 
\]
Accordingly, we can conclude that the shift vector $\beta^{(3)i}_T$ is
of the second order  and obtain
the following equation at the second order $\lambda^2$
of Eq.(\ref{eqn:shiftdiffer}): 
\begin{eqnarray}
\Delta_X\beta^{(3)i}_T=\lambda^2 
\frac{4}{t_I^3(1+z_I)^2}\left(\frac{t}{t_I}\right)^{-1/3}
\left(-\Psi^{(1)}_{N|jj}\Psi^{(1)}_{N|i}+\Psi^{(1)}_{N|ij}\Psi^{(1)}_{N|j}
+\Psi^{(2)}_{N|i}\right).\label{eqn:shiftpoisson}
\end{eqnarray}
The consistency of our  formulation can be easily checked 
if one confirms  that  the right-hand side of this equation  satisfies     
also divergenceless condition with respect to
the Lagrangian coordinates  by using Eq.(\ref{eqn:newvelsecond}). 
Using Green function of the Laplacian, we can express
the solution of this equation as
\begin{equation}
 \beta^{(3)i}_T(\bm{X},t)= \lambda^2\frac{4}{t_I^3(1+z_I)^2}
  \left(\frac{t}{t_I}\right)^{-1/3}\bar{\beta}_T^{(3)i}(\bm{X})
  +O\!\left(\lambda^3\right) \label{eqn:shiftsol},
 \end{equation}
where
\begin{eqnarray}
\bar{\beta}_T^{(3)i}(\bm{X}):= \int\!
 d^3\!\bm{Y}\frac{\left(\Psi^{(1)}_{N|jj}(\bm{Y})
\Psi^{(1)}_{N|i}(\bm{Y})
  -\Psi^{(1)}_{N|ij}(\bm{Y})\Psi^{(1)}_{N|j}(\bm{Y})
-\Psi^{(2)}_{N|i}(\bm{Y})\right)}{4\pi |\bm{X}-\bm{Y}|}.\label{eqn:shifspatial}
\end{eqnarray}
This is the expression of the shift vector 
that we have looked for in order to solve Eq.(\ref{eqn:pntransdiff}). 
Thus we are able to express all perturbation quantities of the metric 
in our coordinates  in terms of the trajectory field. 
This must be so because it is only a dynamical variable 
in the Lagrangian description.  
%%%%%%%%%%%%%%%%%%%%%%%%%%%%%%%%%%%%%%%%%%%%%%%%%%%%%%%%%%%%%%%%%%%%%
If noting that we have considered only the shift vector $\beta^i$ not
$\beta_i$, we find that the second order solution of the metric 
component $g_{0i}$ has a growing mode: $g^{(3)}_{0i}=a^2\beta^i\propto
t$. 
This time dependence exactly agrees with the results derived recently 
by Matarrese, Mollerach \& Bruni \shortcite{mpphy}. (The comparison 
needs the caution that they used the conformal coordinates.) 
However, 
our approach is essentially different from them:  
they derived the second order solutions of 
metric perturbations in the Poisson gauge, which reduces to the
longitudinal gauge at the lowest order with respect to the scalar mode, 
by transforming the known solutions in the synchronous comoving
coordinates according to  the second order gauge transformation. 
Similarly, it will be easy to check whether 
other second order solutions of metric perturbations 
 derived by our formalism agree with 
their results. Here, we are interested in the PN solution of the
trajectory field. 
%%%%%%%%%%%%%%%%%%%%%%%%%%%%%%%%%%%%%%%%%%%%%%%%%%%%%%%%%%%%%%%%%%%%

By inserting Eq.(\ref{eqn:shiftsol}) into Eq.(\ref{eqn:pntransdiff}),
we can obtain the following
equation at the lowest order of $\lambda$:
\begin{equation}
 \left(\frac{d^2}{dt^2}+2\frac{\dot{a}}{a}\frac{d}{dt}\right)\nabla_X\times
  {}^{(PN)}\!\bm{S}^{(1)T}(\bm{X},t)=\mbox{{\bf 0}}.
\end{equation}
As an irrotational  case, the form of this equation  allows us to seek the
solutions of the following form:
\[
 {}^{(PN)}\!S^{(1)T}_i(\bm{X},t)=Q^T_z(t){}^{(PN)}\!\Xi_i^{(1)}(\bm{X}),
\hspace{3em}
Q^T_z(t_I)=\nabla_X\cdot{}^{(PN)}\!\bm{\Xi}^{(1)}=0,   
\]  
and from the above equations we can obtain
\begin{equation}
 \ddot{Q}^T_z+2\frac{\dot{a}}{a}\dot{Q}^T_z=0, \label{eqn:pntransfirst}
\end{equation}
with
\begin{equation}
 {}^{(PN)}\!\bm{\Xi}^{(1)}(\bm{X})=\bm{\Pi}(\bm{X}),\hspace{2em}\nabla_X\cdot\bm{\Pi}=0,
\end{equation}
where $\bm{\Pi}$ is the unknown function determined by the initial
conditions. The solutions of equation (\ref{eqn:pntransfirst}) 
can be easily found to have  only a  decaying mode. The first
order solution  may be safely
 ignored because the decaying mode play physically no  important role:
\begin{equation}
 Q^T_z{}^{(PN)}\Xi^{(1)}_i\approx 0.\label{eqn:pntransfirstsol}
\end{equation}
Next, we similarly consider the second order solution of 
the PN transverse part with the form
\begin{equation}
 {}^{(PN)}\!S^{(2)T}_i(\bm{X},t)=Q^T_{zz}(t){}^{(PN)}\Xi_i^{(2)}(\bm{X}),
  \hspace{2em}Q^T_{zz}(t_I)=
  \nabla_X\cdot{}^{(PN)}\!\bm{\Xi}^{(2)}=0.
\end{equation}
Inserting this ansatz, Eqs.(\ref{eqn:shiftsol}) and
(\ref{eqn:pntransfirstsol}) 
into Eq.(\ref{eqn:pntransdiff}),
we can obtain the following equation at
the second order in $\lambda$:
\begin{equation}
 \left(\ddot{Q}^T_{zz}+2\frac{\dot{a}}{a}\dot{Q}^T_{zz}\right)
   \nabla_X\times {}^{(PN)}\!\bm{\Xi}^{(2)}=-\frac{4}{t_I^4(1+z_I)^2}
 \left(\frac{t}{t_I}\right)^{-4/3}\nabla_X \times\bar{\bm{\beta}}^{(3)}_T.
\label{eqn:pnvortdiff}
\end{equation}
The solution of this equation can be found by solving
the linear ordinary differential equations:
\begin{equation}
 \ddot{Q}^T_{zz}+2\frac{\dot{a}}{a}\dot{Q}^T_{zz}=\frac{4}{t_I^2}
  \left(\frac{t}{t_I}\right)^{-4/3},\label{eqn:transsecond}
\end{equation}
with
\begin{eqnarray}
 \Delta_X{}^{(PN)}\!\Xi^{(2)}_i&=&-\frac{1}{(1+z_I)^2t_I^2}\Delta_X
  \bar{\beta}^{(3)i}_T \nonumber \\
  &=&-\frac{1}{(1+z_I)^2t_I^2}\left(-\Psi^{(1)}_{N|jj}\Psi^{(1)}_{N|i}
 +\Psi^{(1)}_{N|ij}\Psi^{(1)}_{N|j}+\Psi^{(2)}_{N|i}\right).
\end{eqnarray}
Since we expect the quantity $Q^T_{zz}{}^{(PN)}\!\Xi^{(2)}_i$ to be much 
smaller than the order $\lambda$ quantities such as $\mal{\delta}$ at
the initial time,  
we can probably adopt again the following initial condition: 
\begin{equation}
 \dot{Q}^T_{zz}(t_I){}^{(PN)}\!\Xi^{(2)}_i=0.
\end{equation}
Then we can find the solution of Eq.(\ref{eqn:transsecond}):
\begin{eqnarray}
 Q^T_{zz}=\left(\frac{3}{2}\right)\left[4\left(\frac{t}{t_I}\right)^{2/3}-12
 +8\left(\frac{t}{t_I}\right)^{-1/3}\right].
\end{eqnarray}
Thus the transverse part of the first PN displacement vector has a growing
mode
even if the Newtonian trajectory field does not have a vorticity flow.
We also  note that the solution with the growing mode is a  particular
solution of the differential equation (\ref{eqn:pnvortdiff}). 
From the observational viewpoint, 
it will be important that we was able to 
 connect the solutions of vector perturbations 
 with the initial density fluctuation 
field which is available through observations of 
the anisotropies of CMB
and the two-point correlation function of galaxies. 

\section{Results and Implications}

Summarizing the results derived  until the previous section, we have obtained
the following family of trajectories as a  physical complete PN solution
up to the second order of $\lambda$ superposed on an
isotropic homogeneous deformation solution of Einstein-de
Sitter background:
{
\setcounter{enumi}{\value{equation}}
\addtocounter{enumi}{1}
\setcounter{equation}{0}
\renewcommand{\theequation}{\thesection.\theenumi\alph{equation}}
\begin{eqnarray}
&&\bm{f}(\bm{X},t)
=\bm{X}
+q_z(t)  \nabla_X \Psi^{(1)}_N(\bm{X}) 
+q_{zz}(t)  \nabla_X \Psi^{(2)}_N(\bm{X})
+\frac{1}{c^2}Q_{zz}^{(b)}(t)\nabla_X\Psi_{PN}(\bm{X})
 +\frac{1}{c^2}Q^T_{zz}(t){}^{(PN)}\!\bm{\Xi}^{(2)}(\bm{X}),
\label{eqn:completetraj}
\end{eqnarray}
with 
\begin{eqnarray}
&&q_z=\left(\frac{3}{2}\right)\left[\left(\frac{t}{t_I}\right)^{2/3}-1\right],\\
&&q_{zz}=\left(\frac{3}{2}\right)^2\left[-\frac{3}{14}\left(\frac{t}{t_I}\right)^{4/3}
+\frac{3}{5}\left(\frac{t}{t_I}\right)^{2/3}-\frac{1}{2}+\frac{4}{35}\left(
\frac{t}{t_I}\right)^{-1}\right],\\
&&Q_{zz}^{(b)}=\left(\frac{3}{2}\right)\left[\frac{2}{5}\left(\frac{t}{t_I}\right)^{2/3}
-\frac{2}{3}+\frac{4}{15}\left(\frac{t}{t_I}\right)^{-1}\right], \\
&&Q^T_{zz}=\left(\frac{3}{2}\right)\left[4\left(\frac{t}{t_I}\right)^{2/3}-12
 +8\left(\frac{t}{t_I}\right)^{-1/3}\right],
\end{eqnarray}
and
\begin{eqnarray}
&&\Delta_X\Psi^{(1)}_N=-\frac{2}{3}\mal{\delta}_g,\\
&&\Delta_X\Psi^{(2)}_N=\Psi^{(1)}_{N|ii}\Psi^{(1)}_{N|jj}
-\Psi^{(1)}_{N|ij}\Psi^{(1)}_{N|ij},\\
&&\Delta_X\Psi_{PN}=\frac{1}{(1+z_I)^2t_I^2}\!\!
\left[2\mal{\delta}_g\Psi^{(1)}_N\!-\!\frac{1}{2}\left(\Psi^{(1)}_{N|i}\right)^2
+\!\frac{15}{2}
\int\!d\bm{Y}\frac{\left(\Psi^{(1)}_{N|ij}(\bm{Y})\Psi^{(1)}_{N|i}(\bm{Y})
\right)_{|j}}{4\pi
 |\bm{X}-\bm{Y}|}\!+\!\frac{9}{2}\Psi_N^{(2)}\right],\!\\
&&\Delta_X{}^{(PN)}\!\Xi^{(2)}_i
  =-\frac{1}{(1+z_I)^2t_I^2}\left(-\Psi^{(1)}_{N|jj}\Psi^{(1)}_{N|i}
 +\Psi^{(1)}_{N|ij}\Psi^{(1)}_{N|j}+\Psi^{(2)}_{N|i}\right).
\end{eqnarray}\setcounter{equation}{\value{enumi}}}\noindent
%

%%%%%%%%%%%%%%%%%%%%%%%%%%%%%%%%%%%%%%%%%%%%%%%%%%%%%%%%
We remark that  both  longitudinal and transverse
perturbation solutions 
of the first PN displacement vector have a growing mode
and the same time dependence, namely, $\propto t^{2/3}$.
In particular, the appearance of the transverse part in 
PN displacement vector is interesting because there is no correspondence 
in Newtonian order up to the third order in $\lambda$ \cite{buchert94}.
%%%%%%%%%%%%%%%%%%%%%%%%%%%%%%%%%%%%%%%%%%%%%%%%%%%%%%%%%%%%%%%%%%%%%%%%% 
The physical interpretation is examined by us 
\shortcite{tfvort1} in detail. 
According to the study, it has been shown that 
the existence of the growing transverse mode guarantees 
the PN conservation law of vorticity field along flow lines 
derived by extending the Newtonian  
{\it Kelvin's circulation theorem} for the collisionless fluid. 
%%%%%%%%%%%%%%%%%%%%%%%%%%%%%%%%%%%%%%%%%%%%%%%%%%%%%%%%%%%%%%%%%%%%%%%%%%

%%%%%%%%%%%%%%%%%%%%%%%%%%%%%%%%%%%%%%%%%%%%%%%%%%%%%%%%%%%%%%%%%%%%%%
To see the effect of the PN corrections clearly, let us make a simple order
estimation.  For the purpose, 
 consider the initial density fluctuation
$\mal{\delta}_{g(l)}$ with characteristic length $l$ in the comoving frame.
It should be noted that, due to
$a(t_0)=1$, we may regard the length $l$ as the physical scale at the
present time $t_0$. Therefore, if, for convenience, 
 we consider $\dot{f}^i(\bm{X},t)$ which
represents the velocity field 
of the fluid elements in the comoving frame, it
may be estimated roughly at some time $t$ as follows: 
\begin{eqnarray}
&&\dot{f}^i_{(l)}(t)\sim l\left[\dot{q}_z \mal{\delta}_{g(l)}
+\dot{q}_{zz} \left(\mal{\delta}_{g(l)}\right)^2\right] 
+l\dot{Q}^{(b)}_{zz}\left(\frac{l}{(1+z_I)ct_I}\right)^2\left(
\mal{\delta}_{g(l)}\right)^2
+l\dot{Q}^T_{zz}\left(\frac{l}{(1+z_I)ct_I}\right)^2\left(
\mal{\delta}_{g(l)}\right)^2+O\!\left(\lambda^3,c^{-4}\right),
\end{eqnarray}
where we used the following estimation:
\begin{eqnarray}
&&\Psi^{(1)}_{N(l)}\sim \mal{\delta}_{g(l)}l^2,\nonumber \\
&&\Psi^{(2)}_{N(l)}\sim \left(\mal{\delta}_{g(l)}\right)^2l^2, \nonumber \\ 
&&\Psi_{PN}\sim \left(\frac{l}{(1+z_I)t_I} \right)^2\left(\mal{\delta}_{g(l)}
\right)^2l^2, \nonumber \\
&& {}^{(PN)}\!\Xi^{(2)}_i\sim
 \left(\frac{l}{(1+z_I)t_I} \right)^2\left(\mal{\delta}_{g(l)}
\right)^2l^2.
\end{eqnarray}
Noting  $ct_I=ct_0/(1+z_I)^{3/2}$, 
$q_z\simeq (1+z_I)a$ and so on, the above equation becomes
\begin{equation}
\dot{f}^i_{(l)}(t)\sim l\left[ \dot{a} \delta_{g(l)}(t_0)+\dot{a}a
\left(\delta_{g(l)}(t_0)\right)^2\right] 
+l\dot{a}\left(\frac{l}{ct_0}\right)^2\left(\delta_{g(l)}(t_0)
\right)^2+\cdots, \label{eqn:aab}
\end{equation}
where $\delta_{g(l)}(t_0)$ is defined by $(1+z_I)\mal{\delta}_{g(l)}$ 
representing the present value of the density contrast for scale $l$
extrapolated  by the linear theory. Note that both the longitudinal  
and the transverse parts of the PN displacement vector are
of the same order of magnitude. 

This expression suggests the following interpretations. 
First, we can conclude that the first and second terms 
in the right-hand side of Eq.(\ref{eqn:aab}) agree with the result 
calculated by Newtonian theory if they are expressed in the terms of
the gauge invariant quantities at the initial time, 
and the third  term is the first nonvanishing PN correction whose order
is $(l/(ct_0))^2(\delta_{g(l)}(t_0))^2$. 
Note that $ct_0$ is the horizon scale at present time $t_0$. 
Comparing the first term with the third term
on the right-hand side, it is also obvious that
the  expansion parameter of  our scheme 
is the
perturbation quantity 
$(l/(ct_0))^2\delta_{g(l)}(t_0)$. 
Consequently, 
the above expression for the field of trajectories 
may describe only the congruence of the fluid elements inside 
the scale satisfying the following condition:
\begin{equation}
l\ll
 \frac{ct_0}{\left(\delta_{g(l)}(t_0)\right)^{1/2} }.
\end{equation}
If applying the usual power spectrum to the above density fluctuation 
field, 
this condition means that the solution (\ref{eqn:aab}) can describe
the motion of fluid elements up to quasi nonlinear regime
for fluctuations 
not only inside but also beyond the {\it present} Hubble
horizon scale $ct_0$, because the magnitude of density contrast 
at the scale $ct_0$ is much smaller than unity. 
The COBE microwave background measurement suggests that
the power spectrum has the positive slopes on the large scale, 
supporting the assumption of horizon-scale homogeneity on the horizon
scale.    
Furthermore,  Eq.(\ref{eqn:aab}) means that as long as we consider 
the evolution of fluctuation field whose characteristic scale $l$ is 
much smaller than the horizon scale $ct_0$, the fluctuation is well 
described by  Newtonian theory because Newtonian correction 
term, namely the second term in Eq.(\ref{eqn:aab}), 
 keeps being much larger than 
the first PN correction term from the initial time 
until today. 
This may not be correct when one consider the fluctuations with larger scale. 
In fact we can evaluate the redshift  $z_N$ when the Newtonian correction 
term
 begins to become larger than the PN correction
term for the density fluctuation with the scale $l$:
\begin{equation}
1+z_N\sim \left(\frac{ct_0}{l}\right)^2.
\end{equation}
This redshift is about the same as the redshift when the fluctuation 
of the scale $l$ just entered into the Hubble horizon $ct_N$.
In other words, the PN correction might play an important role 
for the early evolution of the density contrast.
  For example,  
SDSS is expected to survey the region of several hundred megaparsecs cube, 
then the above estimate gives $z_N \sim 40$ for the
fluctuation with the scale $300h^{-1}\mbox{Mpc}$.
Since the density field at 
such  a redshift is still in the linear stage, 
the PN effects may not be so important for such scales.  
If this is the case, it will confirm the validity of
the use of Newtonian simulation for such scales. However,  
it should be noted that we could come to these conclusions 
by formulating 
the Lagrangian perturbation theory in the longitudinal gauge and
expressing the solution of the trajectory field 
in terms of the gauge invariant quantities at the initial time. 
These conclusions  may be clearer if one recalls that set
of equations in the relativistic linearized  theory can be reduced to 
a Newtonian-like system of equations expressed 
in terms of the gauge invariant 
variables in the longitudinal gauge (see Appendix A).  
Furthermore, this gauge condition allows us to have a straightforward 
Newtonian limit on scales much smaller than the horizon scale even in
the nonlinear situations. 
The PN corrections are small compared with the Newtonian solutions. 
However, because the first PN solutions has a  transverse part with
a growing mode which  cannot be predicted by   
Newtonian theory  as well as by   gauge invariant linearized theory,
its characteristic effects might turn out on the large scales.
Actually, if we compared the order of PN transverse part with 
Newtonian transverse one at third order of $\lambda$ derived by
Buchert \shortcite{buchert94},  
we can estimate 
the redshift $z_{(T)N}$ when the two become the same order of 
magnitude  
for the fluctuation with  the scale $50h^{-1}\mbox{Mpc}$ as 
\begin{equation}
1+z_{(T)_N}\sim 4.
\end{equation}
This shows that the PN transverse mode might play an important role in
the observational cosmology. 
In particular, the effects may induce the secondary temperature anisotropies 
of CMB  as the Rees-Sciama effect \shortcite{rees} and potentially be observable.  
%%%%%%%%%%%%%%%%%%%%%%%%%%%%%%%%%%%%%%%%%%%%%%%%%%%%%%%%%%%%%%%%%%%%%%%%%%%%%%
Mollerach \& Matarrese \shortcite{mollerach} investigated 
the secondary CMB anisotropies induced by the second order metric perturbations 
during the travel of CMB photons
 from the last scattering surface to us, using the formalism developed 
by Pyne and Carroll \shortcite{pynecar}. 
They concluded that the effect of the higher order perturbations 
is much smaller than that of the usual gravitational lensing due to 
 the first
order scalar perturbation. 
However, the study of second order anisotropies will be relevant 
because they have a possibility to produce a non negligible contribution 
compared  to the first order ones, due to the long distances of the
travel. The reason is that, since the second order effects are caused by 
the integrals of the non-linearities in the metric 
 perturbations along the photon path, 
they  accumulate the small effects.
 Moreover, the second order effects 
 do not produce
 the cancellation for the integral along  
photons path  in contrast with the linear effects, so 
they may give the primary contribution to some statistical measures as, 
for example, the three-point function of temperature anisotropies. 
Thus, it is important 
to estimate the magnitude of 
anisotropies due to  
second order vector 
perturbations as a possible contribution to the linear
order anisotropy calculations.  
It should be emphasized that our formalism allows us to estimate the magnitude 
of the second order vector mode 
of metric perturbations, namely 
 the shift vector,  by using the power spectrum of density fluctuation field.     
We are now investigating the secondary effect using the  
``{\it total angular momentum method}'' developed by 
Hu \& White \shortcite{Hu}. 
According to the formalism, it seems that the growing 
vector perturbation produces a behavior of anisotropy 
 largely different form the scalar mode results. 
It will be interesting to compare our
result with that of Mollerach \& Matarrese. 
%%%%%%%%%%%%%%%%%%%%%%%%%%%%%%%%%%%%%%%%%%%%%%%%%%%%%%%%%%%%%%%%%%%%%%%%%%%
Furthermore, for the N-body simulations of more wide region of the universe, 
the PN corrections might also play important role. 
Our approach has an advantage to be used in numerical work based on the
presently available Newtonian simulation.
In any case, it would be very interesting to see these
effects due to the PN corrections.
These  works are  now in progress. 

%%%%%%%%%%%%%%%%%%%%%%%%%%%%%%%%%%%%%%%%%%%%%%%%%%%%%%%%%%%%%%%%%%%%%%%%%%

\section{Summary}

In this paper, we have formulated the PN Lagrangian perturbation
theory in the perturbed Einstein-de Sitter universe using
the (3+1) formalism  
and investigated their effects on the evolution of the
large-scale structure of the universe up to the quasi nonlinear regime.
In this formalism, all force terms in the first PN equations of
motion are expressed in terms of the (gauge invariant) 
Newtonian quantities. 
As a result, we can solve the trajectory field iteratively
in the PN approximation. 

Our formulation based on the gauge condition (\ref{eqn:a}) has 
a natural Newtonian limit and, consequently,  we can show 
that the longitudinal part of shift vector $\beta_L^{(3)}$ and 
slice condition $\theta$, which are freely specified 
within the residual gauge freedom, do not influence the 
solution of the trajectory field only 
if the Newtonian one is assumed  to be
irrotational.
 
It has  been shown that the longitudinal  solutions 
of the first PN displacement vector in our coordinate conditions
have  spurious modes caused by the gauge ambiguities 
and, if  the solution is expressed in the
terms of the gauge invariant quantities at the initial time, 
the PN irrotational lowest order solution (order $\lambda$) take exactly same 
expression with corresponding Newtonian first order solution. 
This allows us to regard the gauge invariant lowest order solution 
as a physical Newtonian lowest order one.
On the other hand, since  the transverse solution of the PN displacement vector 
is generated from the beginning 
by the gauge invariant quantity, namely, the transverse
part of the shift vector  $\beta_T^{(3)i}$, it  first appears with  the same 
order of 
the first non-vanishing PN irrotational solution (order $\lambda^2$)
 expressed in
the terms of the gauge invariant quantities at the initial time. 
Then it was shown
that  the
perturbation quantity $(l/(ct_0))^2\delta_g(t_0)$ become an essential
expansion parameter in the PN approximation in the case that we consider
 the evolution of the fluctuations in the expanding universe.  
The smallness of parameter
  will explain why Newtonian treatment in our coordinate choice 
is so good approximation even for horizon scale fluctuations.
This consideration  seems to be always the case if we study 
the behavior of nonrelativistic matter in the coordinate which has a
straightforward Newtonian limit. 

We have explicitly derived physical nonvanishing PN corrections 
to the Newtonian dynamics and found some interesting results. 
One is that the first PN displacement vector has a transverse part with
a growing mode, which  is not  present  in cosmological Newtonian
theory as well as  in the gauge invariant linearized theory.
The effect may be  potentially  observed  
 as a characteristic pattern
of the large-scale structure.
The other is that we was able to estimate rigorously the PN corrections 
to the Newtonian approximation in the situation
we are interested in. 
 
Since the observable region of the large-scale structure increases
steadily, we expect that our work will play some important roles in the
future. 

\section*{Acknowledgment}
We thank  Thomas Buchert and Dr. Arno Wei\ss\
for fruitful discussions and valuable
comments and suggestions. 

\appendix
\section{Gauge Invariant linearized theory}
\setcounter{equation}{0}
\setcounter{section}{1}

In this appendix, we consider the linearized  theory in the
Einstein-de Sitter universe(i.e., ${\cal K}=\Lambda=0$) to determine the
behavior of the metric quantities and the matter variables in the early
stage of the evolution of the density fluctuation. In the linearized
theory,
all perturbation quantities
can be divided into the scalar parts, the vector parts and tensor parts
with respect to the background geometry:
\begin{eqnarray}
&&\beta^i\equiv \beta_{L,i}+\beta_T^i, \nonumber \\
&&v^i\equiv v_{L,i}+v_T^i, \nonumber \\
&&\vdots
\end{eqnarray}
where
\[
\beta_T^i{}_{,i}=v_T^i{}_{,i}=0, \cdots , 
\]
and it should be noted that
we consider only $\beta^i$ 
not $\beta_i$. 
The equations for scalar modes under our
gauge conditions  (\ref{eqn:a})  become 
\begin{eqnarray}
&&\frac{1}{c^2}\frac{\partial  v_L}{\partial t}+\frac{2}{c^2}\frac{\dot{a}}{a}
 v_L+\frac{1}{c}\frac{\partial \beta_L}{\partial t}
+\frac{2}{c}\frac{\dot{a}}{a}\beta_L=-\frac{1}{a^2}\tilde{\alpha},  \\
&& \frac{\partial \delta}{\partial t}-3\frac{\partial \psi}{\partial t}
+\Delta_f v_L=0,  \\
&&\theta=\frac{3}{c}\frac{\dot{a}}{a}\tilde{\alpha}+\frac{3}{c}\frac{\partial \psi}
{\partial t}+\Delta_f \beta_L, \label{eqn:theta}\\
&& \frac{1}{a^2}\Delta_f\tilde{\alpha}=\frac{4 \pi G \rho_b}{c^2}\left(\delta
+3\tilde{\alpha} \right)-\frac{1}{c}\frac{\partial \theta}{\partial t}-\frac{2}{c}\frac{\dot{a}}
{a}\theta, \\
&& \frac{1}{a^2}\Delta_f\psi =\frac{4\pi G \rho_b}{c^2}\delta
+\frac{1}{c}\frac{\dot{a}}{a}\theta, \\
&& \frac{\partial \psi}{\partial t}+\frac{\dot{a}}{a}\tilde{\alpha}
=-4\pi G\rho_ba^2\left(\frac{1}{c^2}v_L+\frac{1}{c}\beta_L\right),
\end{eqnarray}
where $\tilde{\alpha}=\alpha-1$.
 From the above equations, the equation for
$\delta$ can be derived as
\begin{equation}
\frac{\partial^2 \delta_g}{\partial t^2}+2\frac{\dot{a}}{a}\frac{\partial
\delta_g}{\partial t}-4\pi G\rho_b\delta_g=0. \label{eqn:ld}
\end{equation}
where
\begin{equation}
\delta_g=\delta-3\dot{a}a\left(\frac{v_L}{c^2}+\frac{\beta_L}{c}\right)
\label{eqn:gd}.
\end{equation}
Here, $\delta_g$ denotes a gauge invariant quantity, and from Eq.(\ref{eqn:ld}), we
obtain the two evolution modes of the density fluctuation $\delta_g\propto a$ and
$\propto a^{-3/2}$. It should be remarked that this quantity is the only
physical density fluctuation field.
We can make use of the residual gauge freedom for $\beta_L$:
\begin{equation}
\beta_L=0.\label{eqn:longitudinal}
\end{equation}
Then the above equations become
\begin{eqnarray}
&&\Delta_f\tilde{\alpha}=\frac{4\pi Ga^2\rho_b}{c^2}\delta_g, \label{eqn:pois}\\
&&\Delta_f\psi=\frac{4\pi G a^2\rho_b}{c^2}\delta_g, \label{eqn:sp}\\
&&\tilde{\alpha}=-\frac{1}{c^2}\frac{\partial (a^2v_L)}{\partial t}. \label{eqn:lv}
\end{eqnarray}
The Eqs.(\ref{eqn:pois}) and (\ref{eqn:sp})  entirely agree with the
Newtonian-like  Poisson's
 equation. 
However, since general relativity doesn't depend on the scale, Eq.(\ref{eqn:pois})
 is valid for the density fluctuation of all scale in the linear
regime. It is noted that the structure of set of the above equations
is just like Newtonian theory but not exactly same because
the gauge invariant linearized theory must be  constrained by the condition
$\delta_{g}\ll1$. 
Thus we can conclude that the solutions for the expansion 
\begin{eqnarray}
&&\tilde{\alpha}=\frac{\alpha^{(2)}}{c^2}+\frac{\alpha^{(4)}}{c^4}+\cdots, \nonumber \\
&&\psi=\frac{\psi^{(2)}}{c^2}+\frac{\psi^{(4)}}{c^4}+\cdots, \nonumber
\end{eqnarray}
become 
\begin{eqnarray}
&& \Delta_f\alpha^{(2)}=4\pi G a^2\rho_b\delta_g,  \\
&& \alpha^{(4)}=\alpha^{(6)}=\cdots =0,\\
&&\Delta_f\psi^{(2)}=4\pi G a^2\rho_b\delta_g, \\
&&\psi^{(4)}=\psi^{(6)}=\cdots=0.
\end{eqnarray}
Thus if we consider only the growing mode of $\delta_g$, we immediately find
that the solution for $\alpha^{(2)}$ becomes $\propto a^0$. Then the $\delta$
 under our gauge conditions becomes from Eqs.(\ref{eqn:lv}) and (\ref{eqn:gd})
\begin{equation}
\delta_g=\delta+2\frac{\alpha^{(2)}}{c^2}.
\end{equation}
The $\delta$ has the spurious mode $2\alpha^{(2)}/c^2$. This means
that if we use $\delta$ instead of $\delta_g$, the undesirable mode
will appear. Therefore, in order to see the physical density fluctuation,
we must choose $\delta_g$ as a density fluctuation instead of $\delta$.

Finally, we give a physical meaning of the gauge condition
(\ref{eqn:longitudinal}). 
It is easily seen that under the gauge condition 
the extrinsic curvature which represents the cosmic isotropic expansion 
becomes from Eqs.(\ref{eqn:extrinsic}) and (\ref{eqn:theta})
\begin{eqnarray}
K=-3\frac{1}{a}\frac{d a(\tau)}{d\tau}=-3\frac{1}{a(1+\tilde{\alpha})}\frac{da}{cdt},
\end{eqnarray}
where we used the $\psi$ remains constant from Eq.(\ref{eqn:sp}) in the
linear regime. Thus the gauge condition corresponds to the choice of 
the proper time $\tau$ on the spatial hypersurface.


\begin{thebibliography}{}

\bibitem[\protect \citename{Asada, Shibata \& Futamase}1996]{asf}
 Asada H.,  Shibata M., 
 Futamase T., 1996,  Prog. Theor. Phys.,  96, 81 

\bibitem[\protect \citename{Bardeen}1882]{bardeen} 
Bardeen J., 1980,  Phys. Rev. D, 22, 1882

\bibitem[\protect \citename{Bertschinger}1995]
{bertschinger} Bertschinger E., 1995 in  Cosmology and Large Scale 
Structure, Les Houches Session LX,  Schaeffer R. et al. eds., 
 (Elsevier: Science Press, Amsterdam), pp. 273-348. 
 
\bibitem[\protect \citename{Bertschinger \& Hamilton}1994]{berthamil}
Bertschinger E.,   Hamilton A. J. S., 1994, 
 ApJ,  435, 1 

\bibitem[\protect \citename{Bertschinger \& Hamilton}1994]{bertjain} 
Bertschinger E.,  Jain B., 1994, ApJ,  431, 486

\bibitem[\protect \citename{Bouchet et al.}1992]{bouchet92} 
Bouchet F. R., Juszkiewicz R.,  Colombi S., 
 Pellat R., 1992, ApJ,  394, L5

\bibitem[\protect \citename{Bouchet et al.}1995]{bouchet}  
Bouchet F. R., Colombi S.,  Hivon E.,   Juszkiewicz R., 
1995, A\&A,   296, 575
 
\bibitem[\protect \citename{Buchert}1989]{buchert89} Buchert T., 1989,  A\&A,  223, 9 

\bibitem[\protect \citename{Buchert}1992]{buchert92} 
Buchert T., 1992, MNRAS, 254, 729 

\bibitem[\protect \citename{Buchert}1993]{buchert93} 
Buchert T.,  Ehlers J., 1993,  MNRAS,  264, 375

\bibitem[\protect \citename{Buchert}1994]{buchert94} Buchert T, 1994, MNRAS,  267, 811 

\bibitem[\protect \citename{Buchert, Mellot \& Wei\ss}1994]{buchertW}
 Buchert T,   
Melott A. L.,   Wei\ss {}  A. G., 1994, A\&A,
 288, 349

\bibitem[\protect \citename{Buchert \& Ehlers}1993]{be9394}
 Buchert T.,  Ehlers J., 1993, MNRAS,  264, 375 

\bibitem[\protect \citename{Buchert}1995]{buchertpre} Buchert T., in 
Proc. Int. School of Physics Enrico Fermi, Couse CXXXII, Varenna 1995, 
 ed.  S. Bonometto, J. Primack and  A. Provenzale, (IOS Press,
 Amsterdam), p. 543, (astro-ph/9509005) 

\bibitem[\protect \citename{Catelan}1995]{catelan} 
Catelan P., 1995,  MNRAS, 276, 115 
 
\bibitem[\protect \citename{Chandrasekhar}1965]{ch} Chandrasekhar S., 
1965, ApJ,  142, 1488

\bibitem[\protect \citename{Ehlers}1961]{Ehlers} 
Ehlers J., 1961,  Mainz Academy of Science and Literature
(Steinwer, Wiesbaden), No.11, p.792. 

\bibitem[\protect \citename{Ellis}1971]{Ellis} 
Ellis G. F. R., 1971, in General Relativity and
 Cosmology.,  Sachs R. K. ed., (Academic press, New York), p.104-182. 

\bibitem[\protect \citename{Futamase}1988]{futamase88} 
Futamase T., 1988, Phys. Rev. Lett.,  61, 2175

\bibitem[\protect \citename{Futamase}1989]{futamase89} 
Futamase T., 1989, MNRAS,  237, 187

\bibitem[\protect \citename{Futamase}1992]{futamase92} 
Futamase T., 1992, in Proceedings of Gravitational Astronomy,
 Nakamura T. ed., 340  

\bibitem[\protect \citename{Futamase}1995]{futamaselens}
Futamase T., 1995, Prog. Theor. Phys. Lett., 93, 647 

\bibitem[\protect \citename{Futamase}1996]{futamase96} 
Futamase T., 1996, Phys.Rev. D, 53, 681

\bibitem[\protect \citename{SDSS; e.g., Gunn \& Weinberg}1995]{SDSS}
Gunn J. E.,  Weinberg D. H., 1995, in Wide Field Spectroscopy and 
the Distant Universe, ed. S. J. Maddox \& A. Ara\'gon-Salamanca
(Singapore: World Scientific), 3

\bibitem[\protect \citename{Hu \& White}1997]{Hu}
Hu W., White M., 1997, Phys. Rev. D, 56, 596 

\bibitem[\protect \citename{Hui \& Bertschinger}1996]{hui} Hui L.,  
Bertschinger E., 1996,  ApJ,  471, 1

\bibitem[\protect \citename{Irvine}1965]{i} Irvine W. M., 1965,  Ann. Phys.,  32, 322

\bibitem[\protect \citename{Kasai}1992]{kasai92} 
Kasai M., 1992, Phys. Rev. Lett.,  69, 2330 

\bibitem[\protect \citename{Kasai}1993]{kasai93} Kasai M., 1993, Phys. Rev. D, 47, 3214

\bibitem[\protect \citename{Kasai}1995]{kasai95} Kasai M., 1995, Phys. Rev. D, 52, 5605

\bibitem[\protect \citename{Kofman \& Pogosyan}1995]{kofman} 
Kofman L.,   Pogosyan D., 1995,  ApJ,  442, 30  

\bibitem[\protect \citename{Kodama \& Sasaki}1984]{ks} 
Kodama H., Sasaki M., 1984, Prog. Theor. Phys. Suppl.,  78, 1 

\bibitem[\protect \citename{Landau \& Lifshitz}1980]{landau} 
Landau L. D., Lifshitz E. M., 1980, The Classical
 Theory of Fields. Pergamon Press, Oxford

\bibitem[\protect \citename{Ma \& Bertschinger}1995]{mabertschinger} 
Ma C. P.,   Bertschinger E., 1995, 
 ApJ,  455, 7

%\bibitem[\protect \citename{Matarrese, Pantano, \& Saez}1994]{mps1} 
%Matarrese S., Pantano O.,   Saez D., 1994, 
% Phys. Rev. Lett., 47, 1311 

\bibitem[\protect \citename{Matarrese, Pantano, \& Saez}1994]{mps2} 
Matarrese S., Pantano O.,  Saez D., 1994, MNRAS, 271, 513

\bibitem[\protect \citename{Matarrese}1996]{mt} Matarrese  S.,   
Terranova D., 1996,  MNRAS,  283, 400

\bibitem[\protect \citename{Matarrese, Mollerach, \& Bruni}1998]{mpphy}
Matarrese S., Mollerach S.,  Bruni M., 1998, 
 Phys. Rev. D, 58, 043504,  

\bibitem[\protect \citename{Mellot, Buchert \& Wei\ss}1995]{mbw} 
Melott A. L.,  Buchert T.,   Wei\ss {}  A. G., 1995, A\&A,  294, 345

\bibitem[\protect \citename{Mollerach \& Matarrese}1997]{mollerach}
Mollerach S., Matarrese S.,  1997, Phys. Rev. D, 56, 4494

\bibitem[\protect \citename{Moutarde et al.}1991]{moutarde} 
Moutarde F.,  Alimi J.-M.,   Bouchet F. R.,  Pellat R.,
 Ramani A., 1991, ApJ,   382, 377 

\bibitem[\protect \citename{Nariai \& Ueno}1960]{nu} 
Nariai  H.,  Ueno Y., 1960, Prog. Theor. Phys.,  23, 241

\bibitem[\protect \citename{Peebles}1980]{peebles}  
Peebles P. J. E., 1980,   The Large-Scale Structure of the 
Universe., Princeton Univ. Press, Princeton

\bibitem[\protect \citename{Pyne \& Birkinshaw}1996]{pyne}
Pyne T., Birkinshaw M., Apj, 458, 46 

\bibitem[\protect \citename{Pyne \& Carroll}1996]{pynecar}
Pyne T., Carroll S. M., 1996,  Phys. Rev. D, 53, 2920

\bibitem[\protect \citename{Rees \& Sciama}1968]{rees} 
Rees M.,  Sciama D. W., 1968,  Nature,  217, 511  
 
\bibitem[\protect \citename{Shibata \& Asada}1995]{sa} 
Shibata M., Asada H., 1995, 
Prog. Theor. Phys., 94, 11 

\bibitem[\protect \citename{Schneider, Ehlers \& Falco}1992]{schneider} 
Schneider P.,  Ehlers J.,   Falco E. E., 1992, 
 Gravitational Lenses., Springer-Verlag, Berlin  

\bibitem[\protect \citename{Seljak}1996]{seljak}
Seljak U., 1996, Apj, 463, 1

\bibitem[\protect \citename{Smoot et al.}1992]{smoot} Smoot G. F.  et al., 1992,  ApJ,  396, L1

\bibitem[\protect \citename{Takada \& Futamase}1998]{tfvort1} 
Takada M., Futamase T., 1998, Prog. Theor. Phys., 100, 315

\bibitem[\protect \citename{Takada \& Futamase}1998]{tfvort2}
Takada M., Futamase T., 1998, in preparation. 
 
\bibitem[\protect \citename{Tomita}1967]{tomita67} 
Tomita K., 1967,  Prog. Theor. Phys.,  37, 831

\bibitem[\protect \citename{Tomita}1988]{tomita88} 
Tomita K., 1988, Prog. Theor. Phys., 79, 258

\bibitem[\protect \citename{Tomita}1991]{tomita91} 
Tomita K., 1991, Prog. Theor. Phys.   85, 1041

\bibitem[\protect \citename{Weinberg}1972]{w} 
Weinberg S., 1972,  Gravitation and Cosmology.,  Wiley, NewYork 

\bibitem[\protect \citename{Will}1992]{will} 
Will C. M., 1992,  Theory and Experiment in Gravitational
 Relativity., Cambridge University Press

\bibitem[\protect \citename{Wei\ss,{}Gottl\"ober \& Buchert}1996]{weiss} 
Wei\ss{}
   A. G.,  Gottl\"ober S.,  Buchert T., 1996,  MNRAS, 
 278, 953 

\bibitem[\protect \citename{Zel'dovich}1970]{z}  Zel'dovich Ya. B., 1970,  A\&A,  5, 84

\end{thebibliography}
\end{document}